\documentclass[11pt, a4paper]{article}
\usepackage{amsmath}

\newtheorem{theorem}{Theorem}

\newcommand{\benumerate}{\begin{enumerate}}
\newcommand{\eenumerate}{\end{enumerate}}

\newcommand{\bitemize}{\begin{itemize}}

\newcommand{\eitemize}{\end{itemize}}

\begin{document}

\title{On the classification of discrete   Hirota-type equations in 3D }
\author{E.V. Ferapontov, V.S. Novikov, I. Roustemoglou}
    \date{}
    \maketitle
    \vspace{-7mm}
\begin{center}
Department of Mathematical Sciences \\ Loughborough University \\
Loughborough, Leicestershire LE11 3TU \\ United Kingdom \\[2ex]
e-mails: \\
 { \texttt{E.V.Ferapontov@lboro.ac.uk}}\\
  {\texttt{V.Novikov@lboro.ac.uk}}\\
  { \texttt{I.Roustemoglou@lboro.ac.uk}}\\
\end{center}

\bigskip

\begin{abstract}

In the series of recent publications \cite{FerM, FMN, FNR, HN} we have proposed a novel approach to the classification of integrable differential/difference equations in 3D based on the requirement that  hydrodynamic reductions of the corresponding dispersionless limits are `inherited' by the  dispersive equations. In this paper we extend this to the fully discrete   case.  
Based on the method of deformations of hydrodynamic reductions, we classify discrete 3D  integrable Hirota-type equations  within various particularly interesting subclasses. Our method can be viewed as an alternative to the conventional multi-dimensional consistency approach.

\bigskip


\bigskip

\end{abstract}

\newpage

\section{Introduction}

This paper is based on the observation  that various equivalent forms of the 3D Hirota difference equation \cite{Hirota81} can be obtained as `naive' discretisations of second order quasilinear PDEs,  by simply replacing  partial derivatives $\partial$ by discrete derivatives $\triangle$. Although  this  recipe should by no means preserve the integrability in general, it does apply to a whole range of interesting examples. 
Thus, the dispersionless PDE  
$$
(u_1-u_2)u_{12}+(u_3-u_1)u_{13}+(u_2-u_3)u_{23}=0
$$
gives rise to the lattice KP equation \cite{Date, Nijhoff84, Nijhoff86},
\begin{equation}
(\triangle_1u-\triangle_2u)\triangle_{12}u+
(\triangle_3u-\triangle_1u)\triangle_{13}u+
(\triangle_2u-\triangle_3u)\triangle_{23}u=0.
\label{lKP}
\end{equation}
Similarly, the dispersionless PDE
$$
\partial_1 \left(\ln \frac{u_3}{u_2}\right)+\partial_2 \left( \ln \frac{u_1}{u_3}\right)+\partial_3  \left(\ln \frac{u_2}{u_1}\right)=0
$$
results in the  Schwarzian KP equation \cite{Nijhoff84, DN91, BK98, BK99, KonS02},
\begin{equation}
\triangle_1 \left(\ln \frac{\triangle_3u}{\triangle_2u}\right)+\triangle_2 \left(\ln \frac{\triangle_1u}{\triangle_3u}\right)
+\triangle_3 \left(\ln \frac{\triangle_2u}{\triangle_1u}\right)=0.
\label{sKP}
\end{equation}





\noindent Here $u(x^1, x^3, x^3)$ is a function of three (continuous) variables. We use subscripts for partial derivatives of $u$ with respect to the independent variables $x^i$: $u_i=u_{x^i}, \  u_{ij}=u_{x^ix^j},\ \partial_i=\partial_{x^i}, $ etc.  Forward/backward  $\epsilon$-shifts and discrete derivatives in $x^i$-direction  will be denoted $T_i, \ T_{\bar i}$ and  $\triangle_{ i}, \triangle_{\bar i}$, respectively:
$
\triangle_i=\frac{T_i-1}{\epsilon}, ~
\triangle_{\bar i}=\frac{1-T_{\bar i}}{\epsilon}.
$
We also use multi-index notation for multiple shifts/derivatives:
$
T_{i j}=T_iT_{ j}, ~ \triangle_{i\bar j}=\triangle_i\triangle_{\bar j},
$
etc. 

\medskip

Our first  main result (Theorem 1 of Sect. 3) provides a classification of  integrable conservative  equations of the form
\begin{equation}
\triangle_1 f+\triangle_2 g+\triangle_3 h=0,
\label{cons}
\end{equation}
where $f, g, h$ are functions of $\triangle_1u, \triangle_2u, \triangle_3u$ only. Examples of this type appear as $\triangle$-forms of various  discrete equations  of the KP, Toda and Sine-Gordon type, 
see  Appendix for  examples and references.
The corresponding dispersionless limits are scalar conservation laws of the form
$$
\partial_1f(u_1, u_2, u_3)+\partial_2g(u_1, u_2, u_3)+\partial_3h(u_1, u_2, u_3)=0.
$$
The classification is performed modulo elementary transformations of the form $u\to \alpha u+\alpha_ix^i$, as well as permutations of the independent variables $x^i$, which preserve the class of discrete conservation laws (\ref{cons}). We show that any integrable equation of the form (\ref{cons}) arises as a conservation law of a  certain discrete integrable equation of  octahedron type,
$$
F(T_1u,\ T_2u,\ T_3u,\ T_{12}u,\ T_{13}u,\ T_{23}u)=0,
$$
see \cite{ABS12} for their classification. More precisely, there exist seven cases of integrable octahedron-type equations (note that our equivalence group is different from the group of admissible transformations
utilised in \cite{ABS12}),  each of them possessing exactly three first order linearly independent  conservation laws of the form (\ref{cons}). Let $I, J, K$ denote  their left hand sides. They give rise to a three-parameter family of integrable equations of the form (\ref{cons}), 
$$
\alpha I+\beta J+\gamma K=0,
$$
where $\alpha, \beta, \gamma$ are three arbitrary constants
(see Theorem 1 for a complete list and explicit formulae). Thus, there exist seven three-parameter families of integrable conservation laws of the form (\ref{cons}). Although we do not obtain any  novel 3D discrete integrable equations (in fact, all existing classification results essentially suggest the uniqueness of  Hirota-type equations \cite{ABS12, TW1}),  their conservative forms seem to be new. One of the seven cases mentioned above is the octahedron equation
$$
(T_2\triangle_1u)(T_3\triangle_2u)(T_1\triangle_3u)=(T_2\triangle_3u)(T_3\triangle_1u)(T_1\triangle_2u),
$$
known as the Schwarzian KP equation in its standard form. It possesses three conservation laws
$$
I=\triangle_2\ln \left(1-\frac{\triangle_3u}{\triangle_1u}\right)-\triangle_3\ln \left(\frac{\triangle_2u}{\triangle_1u}-1\right)=0,
$$
$$
J=\triangle_3\ln \left(1-\frac{\triangle_1u}{\triangle_2u}\right)-\triangle_1\ln \left(\frac{\triangle_3u}{\triangle_2u}-1\right)=0,
$$
$$
K=\triangle_1\ln \left(1-\frac{\triangle_2u}{\triangle_3u}\right)-\triangle_2\ln \left(\frac{\triangle_1u}{\triangle_3u}-1\right)=0,
$$
note that their linear combination $I+J+K=0$ coincides with (\ref{sKP}).

\medskip

Our second main result (Theorem 3 of Sect. 4) is the classification of discrete integrable quasilinear  equations of the form
$$
\sum_{i,j=1}^3 f_{ij}\ \triangle_{ij}u=0,
$$
where $f_{ij}$ are functions of $\triangle_1u, \triangle_2u, \triangle_3u$ only. These equations can be viewed as discretisations  of second order quasilinear  PDEs 
$$
\sum_{i,j=1}^3 f_{ij}\ u_{i j}=0
$$
studied in \cite{BFT}. In contrast to the result of Theorem 1, there exists a unique integrable example within this class, namely the lattice KP equation (\ref{lKP}).  

We also classify differential-difference degenerations of  the above equations  with one/two discrete variables
(subsections 3.1, 3.2  and 4.1, 4.2). Some of the examples from Sect. 3.1 are apparently new.

Our approach to the classification of discrete integrable equations is based on the requirement that all hydrodynamic reductions of the corresponding dispersionless limit are inherited by the full discrete (dispersive) equation. This method has been successfully applied recently to  various classes of differential/difference equations in 3D, see \cite{FerM, FMN, FNR, HN}. A brief summary of the method is included in Sect. 2.

In the Appendix we present $\triangle$-forms of various discrete KP/Toda type equations.

\section{Preliminaries: the method of dispersive deformations}

This method applies to dispersive equations possessing a non-degenerate dispersionless limit, and is based on the requirement that all hydrodynamic reductions of the dispersionless limit are `inherited' by the full dispersive (in particular, difference) equation, at least to some finite order in the deformation parameter $\epsilon$, see \cite{FerM, FMN, FNR, HN} for examples and applications. It turns out that all known integrable differential/difference  equations in 3D  satisfy this requirement. Our experience suggests that in most cases it is sufficient to perform calculations up to the order $\epsilon^2$, the necessary conditions for integrability obtained at this stage usually prove to be sufficient, and imply the existence of conventional Lax pairs, etc. 
Let us illustrate our approach by classifying  integrable discrete wave-type equations of the form
\begin{equation}
\triangle_{t \bar t}~u-\triangle_{x \bar x} ~f(u)-\triangle_{y \bar y} ~g(u)=0,
\label{fg}
\end{equation}
where $f$ and $g$ are functions to be determined. Using  expansions of  the form
$$
\triangle_{t \bar t}=\frac{(e^{\epsilon\partial_t}-1)(1-e^{-\epsilon\partial_t})}{\epsilon^2}=\partial_t^2+\frac{\epsilon^2}{12}\partial_t^4+\dots,
$$
we can represent (\ref{fg}) as an infinite series in $\epsilon$, 
$$
u_{tt}-f(u)_{xx}-g(u)_{yy}+\frac{\epsilon^2}{12}[u_{tttt}-f(u)_{xxxx}-g(u)_{yyyy}]+\dots =0.
$$
The corresponding  dispersionless limit  $\epsilon \to 0$ results in the quasilinear wave-type equation
\begin{equation}
u_{tt}-f(u)_{xx}-g(u)_{yy}=0.
\label{fgd}
\end{equation}
This equation possesses exact solutions of the form $u=R(x, y, t)$ where $R$ satisfies a pair of Hopf-type equations
$$
R_y=\mu(R) R_x, ~~~ R_t=\lambda(R) R_x,
$$
and the characteristic speeds $\lambda, \mu$ satisfy the dispersion relation $\lambda^2=f'+g'\mu^2$. Solutions of this type are known as one-phase hydrodynamic reductions, or planar simple waves. Let us require that all such reductions can be deformed into formal solutions of the original equation (\ref{fg}) as follows:
\begin{equation}
 \begin{array}{c}
 R_y=\mu(R) R_x+\epsilon (\dots)+\epsilon^2(\dots)+\dots, \\
\ \\
  R_t=\lambda(R) R_x+\epsilon (\dots)+\epsilon^2(\dots)+\dots,
\end{array}
\label{deform}
\end{equation}
here dots at $\epsilon^k$ denote terms which are polynomial in the $x$-derivatives of $R$ of the order ${k+1}$.
The relation $u=R(x, y, t)$ remains undeformed, this can always be assumed modulo the Miura group. We emphasise that such deformations are required to exist for {\it any} function $\mu(R)$. 
Direct calculation demonstrates that all terms of the order $\epsilon$ vanish identically, while at the order $\epsilon^2$ we get  the following constraints for $f$ and $g$:
$$
f''+g''=0, ~~~ g''(1+f')-g'f''=0, ~~~ f''^2(1+2f')-f'(f'+1)f'''=0.
$$
Without any loss of generality one can set $f(u)=u-\ln (e^u+1), ~ g(u)=\ln (e^u+1)$, resulting in the difference equation
\begin{equation}
\triangle_{t \bar t}~u-\triangle_{x \bar x} ~[u-\ln (e^u+1)]-\triangle_{y \bar y} ~[\ln (e^u+1)]=0,
\label{num1}
\end{equation}
which is yet another equivalent form of the Hirota equation, known as the `gauge-invariant form' \cite{Zabrodin97}, or the `Y-system', see Appendix (we refer to \cite{Kuniba11} for a review of its applications). Its dispersionless limit,
\begin{equation}
u_{tt}-[u-\ln (e^u+1)]_{xx}-[\ln (e^u+1)]_{yy}=0,
\label{num2}
\end{equation}
 appeared recently   in the classification of integrable equations possessing the `central quadric ansatz' \cite{FHZ}. 
 In this case the expansions (\ref{deform}) take the explicit form
$$
 \begin{array}{c}
 R_y=\mu(R)\ R_x+\epsilon^2(a_1R_{xxx}+a_2R_{xx}R_{x}+a_3R_x^3)+O(\epsilon^4), \\
\ \\
  R_t=\lambda(R) \ R_x+\epsilon^2(b_1R_{xxx}+b_2R_{xx}R_{x}+b_3R_x^3)+O(\epsilon^4),
\end{array}
$$ 
where 
\begin{gather*}
\begin{aligned}
a_1 =& \frac{1}{12} \left(\mu ^2-1\right) \mu ', \\
 b_1 =&\frac{\left(\mu ^2-1\right) e^R \left(\mu ^2+2 \mu  \mu ' e^R+2 \mu \mu '-1\right)}{24 \left(e^R+1\right)^2 \lambda },
 \end{aligned}
\end{gather*}
etc. The remaining coefficients $a_i, b_i$ have a far more complicated structure, however, all of them are rational expressions in $\mu$ and its derivatives. They also depend  on $\lambda$ (note that  the dependence on $\lambda$ can  be made linear, as different powers of $\lambda$ can be eliminated via the dispersion relation $\lambda^2=\frac{1}{e^R+1}+\frac{e^R}{e^R+1}\mu^2$).

 \medskip
 
 We emphasise that our approach to the integrability in 3D  is essentially intrinsic: it  applies directly to a given equation, and does not require its embedding into a compatible hierarchy living in a higher dimensional space.

\section {Discrete conservation laws in 3D}

In this section we classify integrable equations of the form (\ref{cons}), 
$$
\triangle_1 f+\triangle_2 g+\triangle_3 h=0,
$$
where $f, g, h$ are functions of $\triangle_1u, \triangle_2u, \triangle_3u$ only.  The corresponding dispersionless limit, 
$$\sum_{i,j=1}^3 f_{ij} (u_k)\ u_{i j}=0,$$
is assumed to be non-degenerate in the following sense:  $\det f_{ij}\ne 0$, and the equation is not linearly degenerate as discussed in \cite{BFT, FMoss}. The classification is performed modulo transformations of the form $u\to \alpha u+\alpha_ix^i$, as well as relabelling of the independent variables $x^i$.

\begin{theorem} Integrable discrete conservation laws are naturally grouped into seven three-parameter families,
$$
\alpha I+\beta J +\gamma K=0,
$$
where $\alpha, \beta, \gamma $ are arbitrary constants, while $I, J, K$ denote left hand sides of three linearly independent discrete conservation laws of the seven octahedron-type  equations listed below. In each case we give explicit forms of $I, J, K$, as well as  the underlying octahedron equation. 

\medskip

\noindent {\bf Case 1.}
   \begin{center}
    \begin{tabular}{  | p{10.3cm} | p{4.8cm} |}
     \hline
     Conservation Laws & Octahedron equation  \\ \hline
     $I=\triangle_1 e^{\triangle_2u} +\triangle_3 \left(e^{\triangle_2u-\triangle_1u}-e^{\triangle_2u} \right)=0$ & $\frac{T_2\tau-T_{12} \tau}{T_{23} \tau}={T_1 \tau}\left(\frac{1}{T_{13} \tau}-\frac{1}{T_3 \tau}\right)$ \\
     $J=\triangle_1 e^{-\triangle_3u} +\triangle_2 \left(e^{\triangle_1u-\triangle_3u}-e^{-\triangle_3u} \right)=0$ & (setting $\tau=e^{u/\epsilon}$) \\
     $K=\triangle_2 \left(\triangle_3u-\ln (1-e^{\triangle_1u}) \right)+\triangle_3 \left(\ln (1-e^{\triangle_1u})-\triangle_1u \right)=0$ &   \\  
      \hline
        \end{tabular}
        \end{center}

\medskip

\noindent {\bf Case 2.}
\begin{center}
    \begin{tabular}{  | p{9.9cm} | p{5.2cm} |}
    \hline
    Conservation Laws & Octahedron equation  \\ \hline
     $I=\triangle_2 \ln \triangle_1u +\triangle_3 \ln \left( 1-\frac{\triangle_2u}{\triangle_1u} \right)=0$ & $T_{12}uT_{13}u+T_{2}uT_{23}u+T_{1}uT_{3}u$ \\
     $J=\triangle_1 \ln \triangle_2u +\triangle_3 \ln \left( \frac{\triangle_1u}{\triangle_2u}-1 \right)=0$ & $=T_{12}uT_{23}u+T_{1}uT_{13}u+T_{2}uT_{3}u$\\
     $K=\triangle_1 \left(\frac{(\triangle_2u)^2}{2}-\triangle_2u \triangle_3u \right)+\triangle_2 \left( \triangle_1u \triangle_3u-\frac{(\triangle_1u)^2}{2} \right)=0$ &  \\     
      \hline 
     \end{tabular}
\end{center}

\medskip

\noindent {\bf Case 3. Generalised lattice Toda} (depending on a parameter $\alpha$)
\begin{center}
    \begin{tabular}{  | p{9.3cm} | p{5.8cm} |}
    \hline
     Conservation Laws & Octahedron equation  \\ \hline
      subcase $\alpha \ne 0$ & \\ 
     $I= \triangle_1 e^{\triangle_2u-\triangle_3u}-\triangle_2 e^{\triangle_1u-\triangle_3u}  =0$ & $\frac{T_{23} \tau}{T_{3} \tau}+\frac{T_{12} \tau}{T_{2} \tau}+\alpha \frac{T_{12} \tau+T_{23} \tau}{T_{2} \tau+T_{3} \tau}=$ \\
    $J= \triangle_2 \ln \left(e^{\triangle_1u}+\alpha \right)+\triangle_3 \left(\ln \frac{e^{\triangle_1u}-e^{\triangle_2u}}{e^{\triangle_1u}+\alpha} -\triangle_2u \right)=0$ & $\frac{T_{12} \tau}{T_{1} \tau}+\frac{T_{13} \tau}{T_{3} \tau}+\alpha \frac{T_{12} \tau+T_{13} \tau}{T_{1} \tau+T_{3} \tau}$\\
    $K= \triangle_1 \ln \left(e^{\triangle_2u}+\alpha \right)+\triangle_3 \left(\ln \frac{e^{\triangle_1u}-e^{\triangle_2u}}{e^{\triangle_2u}+\alpha} -\triangle_1u \right)=0$ &  (setting $\tau=e^{-u/\epsilon}$)  \\ 
      \hline 
    subcase $\alpha = 0$ & lattice Toda equation \\ 
  $I= \triangle_1 e^{\triangle_2u-\triangle_3u}-\triangle_2 e^{\triangle_1u-\triangle_3u}  =0$ & $(T_1-T_3)\frac{T_2\tau}{\tau}=(T_2-T_3)\frac{T_1\tau}{\tau}$ \\
   $J= \triangle_2 \triangle_1u +\triangle_3 (\ln (1-e^{\triangle_2u-\triangle_1u}) -\triangle_2u) =0$ & (setting $\tau=e^{-u/\epsilon}$) \\
   $K= \triangle_1 e^{-\triangle_2u}-\triangle_2 e^{-\triangle_1u}+\triangle_3 (e^{-\triangle_1u}-e^{-\triangle_2u}) =0$ &   \\ 
      \hline 
     \end{tabular}
\end{center}

\medskip

\noindent {\bf Case 4. Lattice KP}
\begin{center}
    \begin{tabular}{  | p{9.1cm} | p{6cm} |}
    \hline
   Conservation Laws & Octahedron equation  \\ \hline
      $I=\triangle_1 ((\triangle_3u)^2-(\triangle_2u)^2)+\triangle_2 ((\triangle_1u)^2$ & $(T_1u-T_2u)T_{12}u+(T_3u-T_1u)T_{13}u$ \\
      $~~~~~~-(\triangle_3u)^2)+ \triangle_3 ((\triangle_2u)^2-(\triangle_1u)^2)=0$ &  $+(T_2u-T_3u)T_{23}u=0$ \\
      $J, K= \alpha_1 \triangle_1 \ln(\triangle_3u-\triangle_2u)+\alpha_2 \triangle_2 \ln(\triangle_1u-\triangle_3u)$ & \\
      $~~~~~~+\alpha_3 \triangle_3 \ln(\triangle_2u-\triangle_1u)=0$ & \\
      where $\alpha_i=const$ and $\alpha_1+\alpha_2+\alpha_3=0$ &   \\
     \hline        
     \end{tabular}
\end{center}

\medskip

\noindent {\bf Case 5. Lattice mKP}
\begin{center}
    \begin{tabular}{  | p{10.3cm} | p{4.8cm} |}
    \hline
    Conservation Laws & Octahedron equation  \\ \hline
     $I=\triangle_1(e^{\triangle_2u}-e^{\triangle_3u})+\triangle_2(e^{\triangle_3u}-e^{\triangle_1u})+\triangle_3(e^{\triangle_1u}-e^{\triangle_2u})=0$ & $\frac{T_{13}\tau-T_{12}\tau}{T_1\tau} +\frac{T_{12}\tau-T_{23}\tau}{T_2\tau}$ \\
     $J=\triangle_1 \ln \left(e^{\triangle_3u}-e^{\triangle_2u} \right)-\triangle_2 \ln \left(e^{\triangle_3u}-e^{\triangle_1u} \right)=0$ & $+\frac{T_{23}\tau-T_{13}\tau}{T_3\tau}=0$ \\
     $K=\triangle_2 \ln \left(e^{\triangle_3u}-e^{\triangle_1u} \right)-\triangle_3 \ln \left(e^{\triangle_2u}-e^{\triangle_1u} \right)=0$ &  (setting $\tau=e^{u/\epsilon}$) \\ 
      \hline        
     \end{tabular}
\end{center}

\medskip

\noindent {\bf Case 6. Schwarzian KP}
\begin{center}
    \begin{tabular}{  | p{10.1cm} | p{5cm} |}
    \hline
    Conservation Laws & Octahedron equation  \\ \hline
    $I=\triangle_2\ln \left(1-\frac{\triangle_3u}{\triangle_1u}\right)-\triangle_3\ln \left(\frac{\triangle_2u}{\triangle_1u}-1\right)=0$ & $(T_2\triangle_1u)(T_3\triangle_2u)(T_1\triangle_3u)$ \\
    $J=\triangle_3\ln \left(1-\frac{\triangle_1u}{\triangle_2u}\right)-\triangle_1\ln \left(\frac{\triangle_3u}{\triangle_2u}-1\right)=0$ & $=(T_2\triangle_3u)(T_3\triangle_1u)(T_1\triangle_2u)$ \\
     $K=\triangle_1\ln \left(1-\frac{\triangle_2u}{\triangle_3u}\right)-\triangle_2\ln \left(\frac{\triangle_1u}{\triangle_3u}-1\right)=0$ &   \\ \hline      
     \end{tabular}
\end{center}

\medskip

\noindent {\bf Case 7. Lattice spin} 
\begin{center}
    \begin{tabular}{  | p{8.8cm} | p{6.3cm} |}
    \hline
    Conservation Laws & Octahedron equation  \\ \hline
     Hyperbolic version  & lattice-spin equation \\  
    $I=\triangle_1\ln \frac{\sinh \triangle_3u}{\sinh \triangle_2u}+\triangle_2 \ln \frac{\sinh \triangle_1u}{\sinh \triangle_3u}+\triangle_3 \ln \frac{\sinh \triangle_2u}{\sinh \triangle_1u}=0$ & $\left(\frac{T_{12}\tau}{T_2\tau}-1  \right)\left(\frac{T_{13}\tau}{T_1\tau}-1  \right)\left(\frac{T_{23}\tau}{T_3\tau}-1  \right)$ \\
      $J=\triangle_1\ln \frac{\sinh (\triangle_2u-\triangle_3u)}{\sinh \triangle_2u}-\triangle_3 \ln \frac{\sinh (\triangle_1u-\triangle_2u)}{\sinh \triangle_2u}=0$ & $=
\left(\frac{T_{12}\tau}{T_1\tau}-1  \right)\left(\frac{T_{13}\tau}{T_3\tau}-1  \right)\left(\frac{T_{23}\tau}{T_2\tau}-1  \right)$ \\
     $K=\triangle_2 \ln \frac{\sinh (\triangle_3u-\triangle_1u)}{\sinh \triangle_1u}-\triangle_3 \ln \frac{\sinh (\triangle_1u-\triangle_2u)}{\sinh \triangle_1u}=0$ & (setting $\tau=e^{2 u/\epsilon}$)  \\ 
      \hline 
     Trigonometric version  & Sine-Gordon equation \\  
  $I=\triangle_1\ln \frac{\sin \triangle_3u}{\sin \triangle_2u}+\triangle_2 \ln \frac{\sin \triangle_1u}{\sin \triangle_3u}+\triangle_3 \ln \frac{\sin \triangle_2u}{\sin \triangle_1u}=0$ & $(T_2\sin \triangle_1u)(T_3\sin \triangle_2u)(T_1\sin \triangle_3u)$ \\
 $J=\triangle_1\ln \frac{\sin (\triangle_2u-\triangle_3u)}{\sin \triangle_2u}-\triangle_3 \ln \frac{\sin (\triangle_1u-\triangle_2u)}{\sin \triangle_2u}=0$ & $=(T_2\sin \triangle_3u)(T_3\sin \triangle_1u)(T_1\sin \triangle_2u)$ \\
     $K=\triangle_2 \ln \frac{\sin (\triangle_3u-\triangle_1u)}{\sin \triangle_1u}-\triangle_3 \ln \frac{\sin (\triangle_1u-\triangle_2u)}{\sin \triangle_1u}=0$ &   \\
      \hline 
     \end{tabular}
\end{center}
\end{theorem}

\medskip

\noindent {\bf Remark.} Although cases 1, 2 do not bare any special name, the corresponding equations can be obtained as degenerations from cases 3-7. Furthermore, they are contained in the classification of \cite{ABS12}.

\centerline{\bf Proof of Theorem 1:}

The dispersionless limit of (\ref{cons}) is a quasilinear conservation law
\begin{equation}
\partial_1 f+\partial_2 g+\partial_3h=0,
\label{fgh1}
\end{equation}
where $f, g, h$ are functions of the variables $a=u_1, \ b=u_2, \ c=u_3$. Requiring that all one-phase reductions of the dispersionless equation (\ref{fgh1}) are inherited by the discrete equation (\ref{cons}) we obtain a set of differential constraints for $f, g, h$, which are the necessary conditions for integrability. Thus, at the order $\epsilon$ we get 
\begin{equation}
f_a=g_b=h_c=0,    ~~~ f_b+g_a+f_c+h_a+g_c+h_b=0.
\label{eps}
\end{equation}
The first set of these relations implies that  the dispersionless limit is equivalent to the second order PDE
\begin{equation}
Fu_{12}+Gu_{13}+Hu_{23}=0,
\label{FGH}
\end{equation}
where
$F=f_b+g_a, \ G=f_c+h_a, \ H=g_c+h_b.$ Note that, by virtue of (\ref{eps}), the coefficients $F, G, H$ satisfy the additional constraint $F+G+H=0$. It follows from \cite{BFT} that, up to a non-zero  factor,  any integrable equation of this type is equivalent to
\begin{equation}
[p(u_1)-q(u_2)]u_{12}+[r(u_3)-p(u_1)]u_{13}+[q(u_2)-r(u_3)]u_{23}=0,
\label{quasi}
\end{equation}
where the functions $p(a), q(b), r(c)$ satisfy the integrability conditions
\begin{equation}
\begin{array}{c}
p''=p'\left( \frac{p'-q'}{p-q}+ \frac{p'-r'}{p-r}- \frac{q'-r'}{q-r}\right),\\
\ \\
q''=q'\left( \frac{q'-p'}{q-p}+ \frac{q'-r'}{q-r}- \frac{p'-r'}{p-r}\right),\\
\ \\
r''=r'\left( \frac{r'-p'}{r-p}+ \frac{r'-q'}{r-q}- \frac{p'-q'}{p-q}\right).
\end{array}
\label{pqr}
\end{equation}
Our further strategy can be summarised as follows:

\begin{itemize}

\item[\bf Step 1.] First, we solve equations (\ref{pqr}). Modulo unessential translations and rescalings this leads to seven quasilinear integrable equations of the form (\ref{quasi}), see the details below.

\item[\bf Step 2.] Next, for all of the seven equations found at step 1, we calculate first order conservation laws. It was demonstrated in \cite{BFT} that any integrable second order quasilinear PDE  possesses exactly four conservation laws  of the form (\ref{fgh1}). 

\item[\bf Step 3.] Taking linear combinations of the four conservation laws in each of the above seven cases, and replacing partial derivatives $u_1, u_2, u_3$ by discrete derivatives $\triangle_1u, \triangle_2u, \triangle_3u$,  we obtain discrete equations (\ref{cons}) which, at this stage, are the {\it candidates} for integrability.

\item[\bf Step 4.] Applying the $\epsilon^2$-integrability test, we obtain constraints for the coefficients of linear combinations. It turns out that only linear combinations of  three (out of four)  conservation laws  pass the  integrability test. In what follows, we present  conservation laws  in such a way that the first three are the ones that pass the integrability test, while the fourth one doesn't. Moreover, each triplet of conservation laws corresponds to one and the same discrete integrable equation of octahedron type. In other words,
there are overall seven discrete integrable equations of octahedron type, each of them possesses three conservation laws, and linear combinations thereof give all integrable examples of the form (\ref{cons}).

\end{itemize}

\medskip

Let us proceed to the solution of the system (\ref{pqr}). There are three essentially different cases to consider,  depending on how many functions among $p, q, r$ are constant (the case when all of them are constant corresponds to linear equations). Some of these cases have additional subcases. These correspond  to the seven cases of Theorem 1, in the same order as they appear below (note that the  labelling below is different, dictated by the logic of the classification procedure).

\medskip

\noindent {\bf Case 1:} $q$ and $r$ are distinct constants. Without any loss of generality one can set $q=1, \ r=-1$. In this case the equations for $q$ and $r$ will be satisfied identically, while the equation for $p$ takes the form
$p''=2pp'^2/(p^2-1)$. Modulo unessential scaling parameters this gives
$
p=(1+e^{u_1})/(1-e^{u_1}),
$
resulting in the PDE
$$
e^{u_1}u_{12}-u_{13}+(1-e^{u_1})u_{23}=0.
$$
This equation possesses four conservation laws:
$$
\partial_1 e^{u_2} +\partial_3 \left(e^{u_2-u_1}-e^{u_2} \right)=0,
$$
$$
\partial_1 e^{-u_3} +\partial_2 \left(e^{u_1-u_3}-e^{-u_3} \right)=0,
$$
$$
\partial_2 \left(u_3-\ln (1-e^{u_1}) \right)+\partial_3 \left(\ln (1-e^{u_1})-u_1 \right)=0,
$$
$$
\partial_1 \left (\frac{u_2 u_3}{2} \right)-\partial_2 \left(\frac{u_1 u_3}{2}-u_1\ln (1-e^{u_1})-Li_2(e^{u_1})\right)+\partial_3 \left(\frac{u_1^2}{2}-\frac{u_1 u_2}{2}-u_1\ln (1-e^{u_1})-Li_2 (e^{u_1}) \right)=0,
$$
where $Li_2$ is the dilogarithm function,  $Li_2(z)=-\int \frac{\ln (1-z)}{z}\ dz$. Applying steps 3 and 4, one can show that discrete versions of the first three conservation laws correspond to  the discrete equation
$$
e^{(T_1u-T_{13}u)/\epsilon}+e^{(T_{12}u-T_{23}u)/\epsilon}=e^{(T_1u-T_{3}u)/\epsilon}+e^{(T_2u-T_{23}u)/\epsilon}.
$$
Setting $\tau=e^{u/\epsilon}$ it can be rewritten as
$$
\frac{T_2\tau-T_{12} \tau}{T_{23} \tau}={T_1 \tau}\left(\frac{1}{T_{13} \tau}-\frac{1}{T_3 \tau}\right).
$$


\medskip

\noindent {\bf Case 2:}  $r$ is constant. Without any loss of generality one can set $r=0$. In this case the above system of ODEs for $p$ and $q$ takes the form
$$
\begin{array}{c}
\frac{p''}{p'}= \frac{p'-q'}{p-q}+ \frac{p'}{p}- \frac{q'}{q}, ~~~ \frac{q''}{q'}= \frac{p'-q'}{p-q}+ \frac{q'}{q}- \frac{p'}{p}.
\end{array}
$$
Subtraction of these equations and the separation of variables leads, modulo unessential rescalings, to the two  different subcases.

\noindent {\bf subcase 2a}:
$p=1/u_1, \ q=1/u_2$. The corresponding PDE is
$$
(u_2-u_1)u_{12}-u_2u_{13}+u_1u_{23}=0.
$$
It possesses four conservation laws:
$$ \partial_2 \ln u_1 +\partial_3 \ln \left( 1-\frac{u_2}{u_1} \right)=0, $$
$$ \partial_1 \ln u_2 +\partial_3 \ln \left( \frac{u_1}{u_2}-1 \right)=0, $$
$$ \partial_1 \left(u_2^2-2 u_2 u_3 \right)+\partial_2 \left(2 u_1 u_3-u_1^2 \right)=0, $$
$$ \partial_1 \left(-\frac{2 u_2^3}{9}+u_2^2 u_3-u_2 u_3^2 \right)+\partial_2 \left(\frac{2 u_1^3}{9}-u_1^2 u_3+u_1 u_3^2 \right)+\partial_3 \left(\frac{u_1^2 u_2-u_1 u_2^2}{3} \right)=0. $$
\noindent Applying steps 3 and 4, one can show that discrete versions of the first three conservation laws correspond to  the discrete equation
$$
T_{12}uT_{13}u+T_{2}uT_{23}u+T_{1}uT_{3}u=T_{12}uT_{23}u+T_{1}uT_{13}u+T_{2}uT_{3}u.
$$

\medskip

\noindent {\bf subcase 2b}: $p=1/(e^{u_1}+\alpha), \ q=1/(e^{u_2}+\alpha)$, $\alpha = const$. The corresponding PDE is
$$
(e^{u_2}-e^{u_1})u_{12}-(e^{u_2}+\alpha)u_{13}+(e^{u_1}+\alpha)u_{23}=0.
$$
If $\alpha \ne 0$ it possesses the following four conservation laws:
$$ \partial_1 (e^{u_2-u_3}+\alpha e^{-u_3}) -\partial_2 (e^{u_1-u_3}+\alpha e^{-u_3}) =0, $$
$$ \partial_2 \ln \left(e^{u_1}+\alpha \right)+\partial_3 \left(\ln \frac{e^{u_1}-e^{u_2}}{e^{u_1}+\alpha} -u_2 \right)=0, $$
$$ \partial_1 \ln \left(e^{u_2}+\alpha \right)+\partial_3 \left(\ln \frac{e^{u_1}-e^{u_2}}{e^{u_2}+\alpha} -u_1 \right)=0, $$
\begin{gather*}
\partial_1 \left(-u_2 u_3+2 u_2 \ln \left(\frac{e^{u_2}+\alpha}{\alpha } \right)+2 Li_2\left(-\frac{e^{u_2}}{\alpha } \right) \right)+
\partial_2 \left(u_1 u_3-2 u_1 \ln \left(\frac{e^{u_1}+\alpha}{\alpha } \right)-2 Li_2\left(-\frac{e^{u_1}}{\alpha } \right) \right) \\
+\partial_3 \left( u_2^2-u_1 u_2+2 \left(u_2-u_1\right) \ln \left(1-e^{u_1-u_2}\right)+2 u_1 \ln \left(\frac{e^{u_1}+\alpha }{\alpha }\right)-2 u_2 \ln \left(\frac{e^{u_2}+\alpha }{\alpha }\right) \right. \\
+\left. 2 Li_2 \left(-\frac{e^{u_1}}{\alpha }\right)-2 Li_2\left(-\frac{e^{u_2}}{\alpha }\right)-2 Li_2\left(e^{u_1-u_2}\right) \right)=0,
\end{gather*}
while when $\alpha=0$ the conservation laws take the form:
$$ \partial_1 e^{u_2-u_3}-\partial_2 e^{u_1-u_3}  =0, $$
$$ \partial_2 u_1 +\partial_3 (\ln (1-e^{u_2-u_1}) -u_2) =0, $$
$$ \partial_1 e^{-u_2}-\partial_2 e^{-u_1}+\partial_3 (e^{-u_1}-e^{-u_2}) =0, $$
$$ \partial_1 (u_2^2-u_2 u_3)+\partial_2 (u_1u_3-u_1^2)+\partial_3 \left(u_1^2-u_1 u_2+ 2 \left(u_2-u_1\right) \ln \left(1-e^{u_1-u_2}\right)-2 Li_2\left(e^{u_1-u_2}\right) \right)   =0.$$
\noindent Applying steps 3 and 4 to the subcase $\alpha \ne 0$, one can show that discrete versions of the first three conservation laws correspond to  the discrete equation
\begin{gather*}
e^{(T_{3}u-T_{23}u) / \epsilon}+e^{(T_{2}u-T_{12}u) / \epsilon}+\alpha e^{(T_{2}u+T_{3}u-T_{12}u-T_{23}u) / \epsilon}= \\
e^{(T_{3}u-T_{13}u) / \epsilon}+e^{(T_1u-T_{12}u) / \epsilon}+\alpha e^{(T_{1}u+T_{3}u-T_{12}u-T_{13}u}.
\end{gather*}
 Setting $\tau=e^{-u/ \epsilon}$, this equation  can be rewritten as
$$
\frac{T_{23} \tau}{T_{3} \tau}+\frac{T_{12} \tau}{T_{2} \tau}+\alpha \frac{T_{12} \tau+T_{23} \tau}{T_{2} \tau+T_{3} \tau}=\frac{T_{12} \tau}{T_{1} \tau}+\frac{T_{13} \tau}{T_{3} \tau}+\alpha \frac{T_{12} \tau+T_{13} \tau}{T_{1} \tau+T_{3} \tau}.
$$
The special case $\alpha=0$ leads to the lattice Toda equation, 
$$
(T_1-T_3)\frac{T_2\tau}{\tau}=(T_2-T_3)\frac{T_1\tau}{\tau},
$$
see  Appendix.

\medskip

\noindent {\bf Case 3:} none of $p, q, r$ are constant. In this case we can separate the variables in (\ref{pqr}) as follows. Dividing equations (\ref{pqr}) by $p', q', r'$ respectively, and adding the first two of them we obtain
$$
p''/p'+q''/q'=2(p'-q')/(p-q).
$$
Multiplying both sides  by $p-q$ and applying the operator  $\partial_a\partial_b$ we obtain
$(p''/p')'=2\alpha p', \  (q''/q')'=2\alpha q',  ~ \alpha=const.$ Thus, $ p''/p'=2\alpha p+\beta_1,\ q''/q'=2\alpha q+\beta_2$.  Substituting these expressions back into the above relation we obtain that $p'$ and $q'$ must be (the same) quadratic polynomials in $p$ and $q$, respectively. Ultimately,
$$
p'=\alpha p^2+\beta p+\gamma, ~~~ q'=\alpha q^2+\beta q+\gamma, ~~~ r'=\alpha r^2+\beta r+\gamma.
$$
Modulo unessential translations and rescalings, this leads to the four subcases.

\noindent {\bf subcase 3a}: $p=u_1, \ q=u_2, \ r=u_3$. The corresponding PDE is
$$
(u_2-u_1)u_{12}+(u_1-u_3)u_{13}+(u_3-u_2)u_{23}=0.
$$
It possesses four conservation laws:
$$ \partial_1 (u_3^2-u_2^2)+\partial_2 (u_1^2-u_3^2)+ \partial_3 (u_2^2-u_1^2)=0, 
$$
$$ \alpha_1 \partial_1 \ln(u_3-u_2)+\alpha_2 \partial_2 \ln(u_1-u_3)+\alpha_3 \partial_3 \ln(u_2-u_1)=0, 
$$
$$ \partial_1 \left(\frac{u_3^3-u_2^3}{3}+\frac{u_2 u_3^2-u_2^2 u_3}{2} \right)+ \partial_2 \left(\frac{u_1^3-u_3^3}{3}+\frac{u_3 u_1^2-u_3^2 u_1}{2} \right)+\partial_3 \left(\frac{u_2^3-u_1^3}{3}+\frac{u_1 u_2^2-u_1^2 u_2}{2} \right)=0, 
$$
where $\alpha_1,\alpha_2,\alpha_3$ are constants satisfying $\alpha_1+\alpha_2+\alpha_3=0$.
\noindent  Applying steps 3 and 4, one can show that discrete versions of the first three conservation laws correspond to  the discrete equation
$$
(T_1u-T_2u)T_{12}u+(T_3u-T_1u)T_{13}u+(T_2u-T_3u)T_{23}u=0,
$$
which is known as the lattice KP equation (see  Appendix).

\medskip

\noindent {\bf subcase 3b}: $p=e^{u_1}, \ q=e^{u_2}, \ r=e^{u_3}$. The corresponding PDE is
$$
(e^{u_1}-e^{u_2})u_{12}+(e^{u_3}-e^{u_1})u_{13}+(e^{u_2}-e^{u_3})u_{23}=0.
$$
It possesses four conservation laws:
$$ \partial_1(e^{u_2}-e^{u_3})+\partial_2(e^{u_3}-e^{u_1})+\partial_3(e^{u_1}-e^{u_2})=0,
$$
$$\partial_1 \ln \left(e^{u_3}-e^{u_2} \right)-\partial_2 \ln \left(e^{u_3}-e^{u_1} \right)=0, 
$$
$$\partial_2 \ln \left(e^{u_3}-e^{u_1} \right)-\partial_3 \ln \left(e^{u_2}-e^{u_1} \right)=0, 
$$
\begin{gather*}
\partial_1 \left(u_2 u_3-u_3^2 +2 \left( u_2-u_3-1\right) \ln \left(1-e^{u_2-u_3}\right)+2 {Li}_2\left(e^{u_2 -u_3}\right) \right) + \\ 
\partial_2 \left( u_3^2-u_1 u_3+2 \left(u_3-u_1+1\right) \ln \left(1-e^{u_1-u_3}\right) -2 Li_2\left(e^{u_1-u_3}\right)      \right) + 
\\ \partial_3 \left( u_1 u_2-u_2^2-2 \left(u_1- u_2\right)+2 \left(u_1- u_2\right) \ln \left(1-e^{u_1- u_2}\right) +2 Li_2\left(e^{u_1-u_2 \text{}}\right)        \right)=0.
\end{gather*}
\noindent  Again, applying steps 3 and 4, one can show that discrete versions of the first three conservation laws correspond to  the discrete equation
$$
e^{-\frac{T_1u}{\epsilon}} (e^{\frac{T_{13}u}{\epsilon}}-e^{\frac{T_{12}u}{\epsilon}})+e^{-\frac{T_2u}{\epsilon}}  (e^{\frac{T_{12}u}{\epsilon}}-e^{\frac{T_{23}u}{\epsilon}})+e^{-\frac{T_3u}{\epsilon}} ( e^{\frac{T_{23}u}{\epsilon}}-e^{\frac{T_{13}u}{\epsilon}})=0.
$$
Setting $\tau=e^{u/ \epsilon}$, this  takes the form 
$$
\frac{T_{13}\tau-T_{12}\tau}{T_1\tau}+\frac{T_{12}\tau-T_{23}\tau}{T_2\tau}+\frac{T_{23}\tau-T_{13}\tau}{T_3\tau}=0,
$$
which is known as the lattice mKP equation (see  Appendix).

\medskip

\noindent {\bf subcase 3c}: $p=1/{u_1}, \ q=1/{u_2}, \ r=1/{u_3}$. The corresponding PDE is
$$
u_3(u_2-u_1)u_{12}+u_2(u_1-u_3)u_{13}+u_1(u_3-u_2)u_{23}=0.
$$
It possesses four conservation laws:
$$
\partial_2\ln \left(1-\frac{u_3}{u_1}\right)-\partial_3\ln \left(\frac{u_2}{u_1}-1\right)=0,
$$
$$
\partial_3\ln \left(1-\frac{u_1}{u_2}\right)-\partial_1\ln \left(\frac{u_3}{u_2}-1\right)=0,
$$
$$
\partial_1\ln \left(1-\frac{u_2}{u_3}\right)-\partial_2\ln \left(\frac{u_1}{u_3}-1\right)=0,
$$
$$
\partial_1 \left(u_2^2 u_3-u_2 u_3^2\right)+\partial_2 \left(u_3^2 u_1-u_3 u_1^2\right)+\partial_3 \left(u_1^2 u_2-u_1 u_2^2\right)=0.
$$
\noindent  Applying steps 3 and 4, one can show that discrete versions of the first three conservation laws correspond to  the discrete equation
$$
(T_2\triangle_1u)(T_3\triangle_2u)(T_1\triangle_3u)=(T_2\triangle_3u)(T_3\triangle_1u)(T_1\triangle_2u),
$$
 known as the Schwarzian KP equation (see  Appendix).

\medskip

\noindent {\bf subcase 3d}: $p=\coth{u_1}, \ q=\coth{u_2}, \ r=\coth{u_3}$ (one can also take the trigonometric version $\coth \to \cot$). The corresponding PDE is
$$
(\coth{u_2}-\coth{u_1})u_{12}+(\coth{u_1}-\coth{u_3})u_{13}+(\coth{u_3}-\coth{u_2})u_{23}=0.
$$
It possesses four conservation laws:
$$
\partial_1\ln \frac{\sinh u_3}{\sinh u_2}+\partial_2 \ln \frac{\sinh u_1}{\sinh u_3}+\partial_3 \ln \frac{\sinh u_2}{\sinh u_1}=0,
$$
$$
\partial_1\ln \frac{\sinh (u_2-u_3)}{\sinh u_2}-\partial_3 \ln \frac{\sinh (u_1-u_2)}{\sinh u_2}=0,
$$
$$ 
\partial_2 \ln \frac{\sinh (u_3-u_1)}{\sinh u_1}-\partial_3 \ln \frac{\sinh (u_1-u_2)}{\sinh u_1}=0,
$$
\begin{gather*}
\partial_1 \left(-2 u_3^2+2 u_2 u_3 -2 u_2 \ln \frac{\sinh (u_2-u_3)}{\sinh u_2}+(2 u_3-1) \ln \frac{\sinh (u_2-u_3)}{\sinh u_3}+  \right.\\
 \left. {Li}_2(e^{2 u_2 })-{Li}_2(e^{2 u_3 })-{Li}_2(e^{2 (u_2 -u_3)})\right)+ \\
\partial_2 \left( 2 u_3^2-2 u_1 u_3+(2 u_1-1) \ln \frac{\sinh (u_3-u_1)}{\sinh u_1}+(1-2 u_3) \ln \frac{\sinh (u_3-u_1)}{\sinh u_3} -      \right. \\
 \left. Li_2(e^{2 u_1})+{Li}_2(e^{2 u_3})+{Li}_2(e^{2 (u_1-u_3)}) \right)+\\
\partial_3 \left( -2 u_2^2+2 u_1 u_2 +2 u_2 \ln \frac{\sinh (u_1-u_2)}{\sinh u_2}+(1-2 u_1) \ln \frac{\sinh (u_1-u_2)}{\sinh u_1}+ \right.\\
 \left. {Li}_2(e^{2 u_1})-{Li}_2(e^{2 u_2 })-{Li}_2(e^{2 (u_1-u_2 )}) \right)=0.
\end{gather*}
\noindent  Applying steps 3 and 4, one can show that discrete versions of the first three conservation laws correspond to  the discrete equation
\begin{gather*}
(e^{2(T_{12}u-T_2u)/ \epsilon}-1) (e^{2(T_{13}u-T_1u)/ \epsilon}-1) (e^{2(T_{23}u-T_3u)/ \epsilon}-1) =  \\
(e^{2(T_{12}u-T_1u)/ \epsilon}-1) (e^{2(T_{13}u-T_3u)/ \epsilon}-1) (e^{2(T_{23}u-T_2u)/ \epsilon}-1).
\end{gather*}
Setting $\tau=e^{2 u/ \epsilon}$, it can be rewritten as
$$
\left(\frac{T_{12}\tau}{T_2\tau}-1  \right)\left(\frac{T_{13}\tau}{T_1\tau}-1  \right)\left(\frac{T_{23}\tau}{T_3\tau}-1  \right)=
\left(\frac{T_{12}\tau}{T_1\tau}-1  \right)\left(\frac{T_{13}\tau}{T_3\tau}-1  \right)\left(\frac{T_{23}\tau}{T_2\tau}-1  \right),
$$
which is known as the lattice spin equation (see  Appendix).
In the trigonometric version, one can show that discrete versions of the conservation laws 
$$
\partial_1\ln \frac{\sin u_3}{\sin u_2}+\partial_2 \ln \frac{\sin u_1}{\sin u_3}+\partial_3 \ln \frac{\sin u_2}{\sin u_1}=0,
$$
$$
\partial_1\ln \frac{\sin (u_2-u_3)}{\sin u_2}-\partial_3 \ln \frac{\sin (u_1-u_2)}{\sin u_2}=0,
$$
$$ 
\partial_2 \ln \frac{\sin (u_3-u_1)}{\sin u_1}-\partial_3 \ln \frac{\sin (u_1-u_2)}{\sin u_1}=0,
$$
correspond to  the discrete Sine-Gordon equation, 
$$
(T_2\sin \triangle_1u)(T_3\sin \triangle_2u)(T_1\sin \triangle_3u)=(T_2\sin \triangle_3u)(T_3\sin \triangle_1u)(T_1\sin \triangle_2u).
$$
This finishes the proof of Theorem 1.

\bigskip

\noindent {\bf Remark.}  It was observed in \cite{Lobb} that the Lagrangians $L(u, u_1, u_2; \alpha_1, \alpha_2)$ of 2D discrete integrable  equations of the ABS type \cite{ABS1} satisfy the closure relations
\begin{equation}
\triangle_1L(u, u_2, u_3; \alpha_2, \alpha_3) + \triangle_2L(u, u_3, u_1; \alpha_3, \alpha_1) + \triangle_3L(u, u_1, u_2; \alpha_1, \alpha_2) = 0,
\label{closure}
\end{equation}
which can be interpreted as   3D discrete conservation laws. For instance, the $Q_1$ case corresponds to the Lagrangian
$$
L(u, u_1, u_2; \alpha_1, \alpha_2)=\alpha_2\ln \left(1-\frac{\triangle_1u}{\triangle_2u}\right)-\alpha_1\ln \left(\frac{\triangle_2u}{\triangle_1u}-1\right).
$$
Remarkably, the corresponding closure relation (\ref{closure}), viewed as a single 3D equation, turns out  to be integrable (subcase 6 of Theorem 1). Note that the constraint $\alpha_1=\alpha_2=\alpha_3$ reduces (\ref{closure}) to the Schwarzian KP equation,
$$
\triangle_1\left(\ln \frac{\triangle_3u}{\triangle_2u}\right)+\triangle_2\left(\ln \frac{\triangle_1u}{\triangle_3u}\right)+\triangle_3\left(\ln \frac{\triangle_2u}{\triangle_1u}\right)=0.
$$
On the contrary, closure relations  corresponding to the Lagrangians containing the dilogarithm $Li_2$ fail the $\epsilon^2$ integrability test. We refer to \cite{Suris} for further connections between  ABS equations and 3D integrable  equations of octahedron type.

\subsection { Two discrete and one continuous variables.}

In this subsection we classify conservative equations of the form
\begin{equation}
\triangle_1 f+\triangle_2 g+\partial_3 h=0,
\label{cons1}
\end{equation}
where $f, g, h$ are functions of $\triangle_1u, \triangle_2u, u_3$. Again, non-degeneracy of the dispersionless limit is assumed. Our classification result is as follows:

\begin{theorem} Integrable equations of the form (\ref{cons1}) are grouped into seven three-parameter families,
$$
\alpha I+\beta J +\gamma K=0,
$$
where $\alpha, \beta, \gamma$ are arbitrary constants, while $I, J, K$ denote left hand sides of three linearly independent semi-discrete conservation laws of the seven differential-difference equations listed below. In each case we give explicit forms of $I, J, K$, as well as the underlying differential-difference equation. 

\medskip

\noindent {\bf Case 1.}
   \begin{center}
    \begin{tabular}{  | p{9.8cm} | p{5.2cm} |}
     \hline
     Conservation Laws & Differential-difference equation  \\ \hline
     $I=\triangle_1 e^{\triangle_2 u} -\partial_3 e^{\triangle_2 u-\triangle_1 u} =0$ & \\
     $J=\triangle_1 u_3 +\triangle_2 \left(e^{\triangle_1 u}-u_3 \right)=0$ &  $\frac{T_{12}v}{T_2v}+\frac{T_{1}v_3}{T_1v}=\frac{T_{1}v}{v}+\frac{T_{2}v_3}{T_2v}$  \\
     $K=\triangle_1 u_3^2+\triangle_2 \left( 2 e^{\triangle_1 u} u_3 -e^{2 \triangle_1 u}-u_3^2\right)-\partial_3 \left( 2 e^{\triangle_1 u} \right)=0$   &  (setting $v=e^{u/ \epsilon}, \partial_3 \to \frac{1}{\epsilon} \partial_3$) \\
     & \\
      \hline
        \end{tabular}
        \end{center}

\medskip

\noindent {\bf Case 2.}
\begin{center}
    \begin{tabular}{  | p{9.5cm} | p{5.6cm} |}
    \hline
    Conservation Laws & Differential-difference equation  \\ \hline
     $I=\triangle_1 (e^{\triangle_2 u}-u_3)+\partial_3 \ln \left(e^{\triangle_1 u}-e^{\triangle_2 u}\right)=0$ &   \\
     $J=\triangle_2 (e^{\triangle_1 u}-u_3)+\partial_3 \ln \left(e^{\triangle_1 u}-e^{\triangle_2 u}\right)=0$ &  $T_{12}v=\frac{T_1vT_2v}{v}+\frac{T_2vT_1v_3-T_1vT_2v_3}{T_2v-T_1v}$  \\
     $K=\triangle_1 (e^{2 \triangle_2 u}- 2e^{\triangle_2 u} u_3+u_3^2)+\triangle_2 (2 e^{\triangle_1 u} u_3-e^{2 \triangle_1 u}-$ &   (setting $v=e^{u/ \epsilon}, \partial_3 \to \frac{1}{\epsilon} \partial_3)$ \\     
		 $~~~~~~u_3^2)+\partial_3 (2 e^{\triangle_2 u}-2 e^{\triangle_1 u})=0$ & \\
      \hline 
     \end{tabular}
\end{center}

\medskip

\noindent {\bf Case 3. }
\begin{center}
    \begin{tabular}{  | p{9.5cm} | p{5.6cm} |}
    \hline
     Conservation Laws & Differential-difference equation  \\ \hline
    $I= \triangle_1 (e^{\triangle_2 u} u_3)-\partial_3 e^{\triangle_2 u} =0$ &   \\
    $J= \triangle_2 (e^{-\triangle_1 u} u_3) +\partial_3 e^{-\triangle_1 u}=0$ &  $\frac{v T_{12}v}{T_1 v}=\frac{T_1 v T_2 v_3}{T_1 v_3}$ (setting $v=e^{u/ \epsilon}$) \\
    $K= \triangle_1 (\triangle_2 u+\ln u_3)-\triangle_2 \ln u_3 =0$ &   \\ 
      \hline 
     \end{tabular}
\end{center}

\medskip

\noindent {\bf Case 4.}
\begin{center}
    \begin{tabular}{  | p{9.5cm} | p{5.6cm} |}
    \hline
   Conservation Laws & Differential-difference equation  \\ \hline
      $I=\triangle_2 \left(\frac{u_3}{\triangle_1 u}\right)-\partial_3 \ln (\triangle_1 u)=0 $ &  \\
      $J=\triangle_1 \ln u_3 +\triangle_2 \ln \left(\frac{\triangle_1 u}{u_3} \right)=0 $ &  $(T_{12}u-T_{2}u)T_1 u_3  =(T_{1}u-u) T_2 u_3$ \\
      $K=\triangle_1 (2u_3 \triangle_2 u )+\partial_3 \left( (\triangle_1 u)^2-2 \triangle_1 u \triangle_2 u   \right) =0 $ &   \\
     \hline        
     \end{tabular}
\end{center}

\medskip

\noindent {\bf Case 5.}
\begin{center}
    \begin{tabular}{  | p{8.9cm} | p{6.2cm} |}
    \hline
    Conservation Laws & Differential-difference equation  \\ \hline
     $I=\triangle_1 (e^{\triangle_2 u}u_3)+\partial_3 (e^{\triangle_2 u-\triangle_1 u}-e^{\triangle_2 u}) =0$ & $v (T_{12}v-T_2 v)T_1 v_3=T_1 v (T_1 v-v)T_2 v_3$ \\
     $J=\triangle_1 \ln u_3+\triangle_2  \ln\left(\frac{1-e^{\triangle_1 u}}{u_3} \right) =0$ &  (setting $v=e^{u/ \epsilon}$)  \\
     $K=\triangle_2 \left( \frac{u_3}{1-e^{\triangle_1 u}}\right) +\partial_3 \left(\ln(1-e^{\triangle_1 u})-\triangle_1 u \right)=0$ &  \\ 
      \hline        
     \end{tabular}
\end{center}

\medskip

\noindent {\bf Case 6. }
\begin{center}
    \begin{tabular}{  | p{9.5cm} | p{5.6cm} |}
    \hline
    Conservation Laws & Differential-difference equation  \\ \hline
    $I=\triangle_1 \ln \left(\frac{\triangle_2 u}{u_3} \right)+\triangle_2 \ln \left(\frac{u_3}{\triangle_1 u} \right) =0$ & $(T_2 \triangle_1u)(\triangle_2u) T_1 u_3  =$ \\
    $J=\triangle_1 \left( \frac{u_3}{\triangle_2 u} \right) +\partial_3  \ln \left(1-\frac{\triangle_1 u}{\triangle_2 u} \right)=0$ & $ (T_1 \triangle_2u)(\triangle_1u) T_2 u_3$ \\
    $K=\triangle_2 \left( \frac{u_3}{\triangle_1 u} \right) +\partial_3  \ln \left(1-\frac{\triangle_2 u}{\triangle_1 u} \right)=0$ &   \\ \hline      
     \end{tabular}
\end{center}

\medskip

\noindent {\bf Case 7. } 
\begin{center}
    \begin{tabular}{  | p{9.5cm} | p{5.6cm} |}
    \hline
    Conservation Laws & Differential-difference equation  \\ \hline
    $I=\triangle_1 \ln \left( \frac{\sinh \triangle_2 u}{u_3} \right) -\triangle_2  \ln \left(\frac{\sinh \triangle_1 u}{u_3} \right)=0$ & $(T_2 \sinh \triangle_1u) (\sinh \triangle_2u) T_1 u_3  =$  \\
    $J=\triangle_1 \left(u_3 \coth \triangle_2 u \right)+\triangle_2 \ln \left(\frac{\sinh(\triangle_1 u-\triangle_2 u)}{\sinh \triangle_2 u} \right) =0$ & $(T_1 \sinh \triangle_2u) (\sinh \triangle_1u) T_2 u_3$ \\
    $K=\triangle_2 \left(u_3 \coth \triangle_1 u \right) +\partial_3  \ln \left(\frac{\sinh(\triangle_1 u-\triangle_2 u)}{\sinh \triangle_1 u} \right)=0$ &  \\ 
      \hline 
     \end{tabular}
\end{center}
\end{theorem}
\noindent {\bf Remark.} See the proof below for  Lax pairs of the above differential-difference equations.

\medskip

\centerline{\bf Proof:}

Our proof is parallel to that of Theorem 1. The dispersionless limit of (\ref{cons1}) is again a quasilinear conservation law of the form (\ref{fgh1}), 
$$
\partial_1 f+\partial_2 g+\partial_3h=0,
$$
where $f, g, h$ are functions of the variables $a=u_1, \ b=u_2, \ c=u_3$. Requiring that all one-phase reductions of the dispersionless equation  are inherited by the discrete equation (\ref{cons1}), we obtain a set of differential constraints for $f, g, h$, which are the necessary conditions for integrability. Thus, at the order $\epsilon$ we get 
\begin{equation}
f_a=g_b=h_c=0,    ~~~ f_c+h_a+g_c+h_b=0,
\label{eps1}
\end{equation}
note the difference with Theorem 1.  The first set of these relations implies that the quasilinear conservation law is equivalent to the second order equation
$$
Fu_{12}+Gu_{13}+Hu_{23}=0,
$$
where
$F=f_b+g_a, \ G=f_c+h_a, \ H=g_c+h_b.$ Note that, by virtue of (\ref{eps1}), the coefficients $F, G, H$ satisfy the additional constraint $G+H=0$. It follows from \cite{BFT} that, up to a non-zero  factor,  any integrable equation of this type is equivalent to
\begin{equation}
[p(u_1)-q(u_2)]u_{12}+r(u_3)u_{13}-r(u_3)u_{23}=0,
\label{quasi1}
\end{equation}
where the functions $p(a), q(b), r(c)$ satisfy the integrability conditions
\begin{equation}
\begin{array}{c}
p''=p'\left( \frac{p'-q'}{p-q}+ (p-q)\frac{r'}{r^2}\right),\\
\ \\
q''=q'\left( \frac{p'-q'}{p-q}- (p-q)\frac{r'}{r^2}\right),\\
\ \\
r''=2\frac{r'^2}{r}.
\end{array}
\label{pqr1}
\end{equation}
Our further strategy is the same as in Theorem 1, namely:
\begin{itemize}
\item[\bf Step 1.] First, we solve equations (\ref{pqr1}). Modulo unessential translations and rescalings this leads to seven quasilinear integrable equations of the form (\ref{quasi1}).

\item[\bf Step 2.] For all of the seven equations found at step 1, we calculate first order conservation laws (there will be four of them in each case).  

\item[\bf Step 3.] Taking linear combinations of the four conservation laws, and replacing $u_1, u_2$ by $\triangle_1u, \triangle_2u$ (keeping $u_3$ as it is), we obtain differential-difference equations (\ref{cons1}) which are the {\it candidates} for integrability.

\item[\bf Step 4.] Applying the $\epsilon^2$-integrability test, we find that only linear combinations of  three conservation laws  (out of four) pass the  integrability test. Below we list conservation laws in such a way that the first three are the ones that pass the integrability test, while the fourth one doesn't. Moreover, each triplet of conservation laws corresponds to one and the same differential-difference equation.
\end{itemize}

Let us begin with the solution of  system (\ref{pqr1}). The analysis leads to seven essentially different cases, which correspond to cases 1-7 of Theorem 2 in the same order as they appear below. First of all, the equation for $r$ implies that there are two essentially different cases: $r=1$ and $r=1/c$.

\medskip

\noindent {\bf Case 1:} $r=1$.  Then equations (\ref{pqr1}) simplify to
$$
p''=p' \frac{p'-q'}{p-q}, ~~~ q''=q' \frac{p'-q'}{p-q}.
$$
There are two subcases depending on how many functions among $p, q$ are constant. 

\medskip

\noindent {\bf subcase 1a:} $q$ is constant (the case $p$=const is similar). Without any loss of generality one can set $q=0$.
 Modulo unessential translations and rescalings this leads to $p=e^a$, 
resulting in the PDE
$$
e^{u_1}u_{12}+u_{13}-u_{23}=0.
$$
This equation possesses four conservation laws:
$$
\partial_1 e^{u_2} -\partial_3 e^{u_2-u_1} =0,
$$
$$
\partial_1 u_3 +\partial_2 \left(e^{u_1}-u_3 \right)=0,
$$
$$
\partial_1 u_3^2+\partial_2 \left( 2 u_3 e^{u_1}-e^{2 u_1}-u_3^2    \right)-\partial_3 \left( 2 e^{u_1} \right)=0,
$$
$$
\partial_1 \left (u_2 u_3 \right)+\partial_2 \left( 2 u_1  e^{u_1}-2 e^{u_1}-u_1 u_3    \right)+\partial_3 \left(  u_1^2-u_1 u_2  \right)=0.
$$
Applying steps 3 and 4, we can show that semi-discrete versions of the first three conservation laws correspond to the differential-difference equation
\begin{equation}
e^{(T_{12}u-T_2u)/ \epsilon}-e^{(T_{1}u-u)/ \epsilon}+T_1 u_3-T_2 u_3=0,
\label{dd1}
\end{equation}
which possesses the Lax pair
$$
T_{2} \psi=e^{(T_1 u-T_2 u)/\epsilon} ~(T_{1}\psi+\psi), ~~~~
\epsilon \psi_3=-e^{(T_{1} u- u)/\epsilon}~( T_{1}\psi+\psi).
$$
Setting $v=e^{u/ \epsilon}$ and $\partial_3 \to \frac{1}{\epsilon} \partial_3$, we can rewrite \eqref{dd1} in the form
$$\frac{T_{12}v}{T_2v}+\frac{T_{1}v_3}{T_1v}=\frac{T_{1}v}{v}+\frac{T_{2}v_3}{T_2v}.$$

\medskip

\noindent {\bf subcase 1b:}  both $p$ and $q$ are non-constant. Modulo unessential translations and rescalings, the elementary separation of variables gives $p=e^a, q=e^b$. The corresponding PDE is
$$
(e^{u_1}-e^{u_2})u_{12}+u_{13}-u_{23}=0.
$$
It possesses four conservation laws:
$$
\partial_1 (e^{u_2}-u_3)+\partial_3 \ln \left(e^{u_1}-e^{u_2}\right)=0,
$$
$$
\partial_2 (e^{u_1}-u_3)+\partial_3 \ln \left(e^{u_1}-e^{u_2}\right)=0,
$$
$$
\partial_1 (e^{2 u_2}- 2e^{u_2} u_3+u_3^2)+\partial_2 (2 e^{u_1} u_3-e^{2 u_1}-u_3^2)+\partial_3 (2 e^{u_2}-2 e^{u_1})=0,
$$
\begin{gather*}
\partial_1 \left(-2 e^{u_2} u_2+u_2 u_3+2 e^{u_2}\right)+\partial_2 \left(2 e^{u_1} u_1-u_1 u_3-2e^{u_1}\right)+ \\
\partial_3 \left(u_1 u_2-u_2^2 +2 \left(u_1-u_2\right) \ln \left(1-e^{u_1-u_2}\right)+2 {Li}_2\left(e^{u_1-u_2}\right)\right)=0.
\end{gather*}
Applying steps 3 and 4, we can show that semi-discrete versions of the first three conservation laws correspond to the differential-difference equation
\begin{equation}
e^{(T_{12}u-T_2u) / \epsilon}-e^{(T_{12}u-T_1u) / \epsilon}+e^{(T_{2}u-u)/ \epsilon}-e^{ (T_{1}u-u) / \epsilon}+T_1 u_3-T_2 u_3=0.
\label{dd2}
\end{equation}
Equation \eqref{dd2} possesses the Lax pair
$$
T_{2} \psi=e^{(T_{1} u-T_{2} u)/\epsilon} ~T_{1}\psi+(1-e^{(T_{1} u-T_{2} u)/\epsilon})\psi, ~~~~
\epsilon \psi_3=e^{(T_{1} u- u)/\epsilon} (T_{1}\psi- \psi).
$$
Note that this case has been recorded before. Setting $v=e^{u/ \epsilon}$ and $\partial_3 \to \frac{1}{\epsilon} \partial_3$, we obtain the equation
$$
T_{12}v=\frac{T_1vT_2v}{v}+\frac{T_2vT_1v_3-T_1vT_2v_3}{T_2v-T_1v},
$$
which has appeared before in the context of  discrete evolutions of plane curves \cite{Adler}. 

\medskip

\noindent {\bf Case 2:}  $ r=1/c$. In this case equations for $p$ and $q$ simplify to
$$
p''=p'\left( \frac{p'-q'}{p-q}- (p-q)\right), ~~~
q''=q'\left( \frac{p'-q'}{p-q}+ (p-q)\right).
$$
There are several subcases depending on how many functions among $p, q$ are constant. 

\medskip

\noindent {\bf subcase 2a}: both $p$ and $q$ are constant. The corresponding PDE is
$$
u_{12}+\frac{1}{u_3}(u_{13}-u_{23})=0.
$$
It possesses four conservation laws:
$$
\partial_1 (e^{u_2} u_3)-\partial_3 e^{u_2} =0,
$$
$$
\partial_2 (e^{-u_1} u_3) +\partial_3 e^{-u_1}=0,
$$
$$
\partial_1 (u_2+\ln u_3)-\partial_2 \ln u_3 =0,
$$
$$
\partial_1 \left (u_2 u_3+2 u_3 \right)+\partial_2 \left( u_1 u_3-2 u_3   \right)-\partial_3 \left(u_1 u_2 \right)=0.
$$
Applying steps 3 and 4, we can show that semi-discrete versions of the first three conservation laws correspond to the differential-difference equation
\begin{equation}
\frac{T_2 u_3}{T_1 u_3}=e^{(T_{12}u-T_{1}u-T_{2}u+u) / \epsilon}.
\label{dd3}
\end{equation}
This equation possesses the Lax pair
$$
T_{1} \psi=-e^{(T_1 u- u)/\epsilon} ~(T_{2}\psi-\psi), ~~~~
\epsilon \psi_3=-u_3 (T_{2}\psi- \psi).
$$
Setting $v=e^{u/ \epsilon}$ we can rewrite \eqref{dd3} as
$$ \frac{v T_{12}v}{T_1 v}=\frac{T_1 v T_2 v_3}{T_1 v_3}.
$$

\medskip

\noindent {\bf subcase 2b}: $q$ is constant  (the case $p$=const is similar). Without any loss of generality one can set $q=0$. The  equation for $p$ takes the form $p''=p'^2/p-pp'$,  which integrates to $p'/p+p=\alpha$. There are further subcases depending on the value of the integration constant $\alpha$.

\medskip

\noindent {\bf subcase 2b(i)}: $\alpha=0$. Then one can take $p=1/a$, which results in the  PDE 
$$
\frac{1}{u_1}u_{12}+\frac{1}{u_3}(u_{13}-u_{23})=0.
$$
It possesses four conservation laws:
$$
\partial_2 (u_3/u_1)-\partial_3 \ln u_1 =0,
$$
$$
\partial_1 \ln u_3 +\partial_2 \ln \left(u_1/u_3 \right)=0,
$$
$$
\partial_1 (2 u_2 u_3)+\partial_3 \left( u_1^2-2 u_1 u_2   \right) =0,
$$
$$
\partial_1 \left (u_2^2 u_3 \right)-\partial_2 \left( \frac{u_1^2 u_3}{3} \right)+\partial_3 \left( u_1^2 u_2-  u_2^2 u_1-\frac{2 u_1^3}{9}\right)=0.
$$
Applying steps 3 and 4, we can show that semi-discrete versions of the first three conservation laws correspond to the differential-difference equation
\begin{equation}
(T_{12}u-T_{2}u)T_1 u_3  =(T_{1}u-u) T_2 u_3 .
\label{dd4}
\end{equation}
This equation possesses the Lax pair
$$
T_{1} \psi=-\frac{(T_1 u- u)}{\epsilon} ~T_{2}\psi+\psi, ~~~~
\epsilon \psi_3=-u_3 T_{2}\psi.
$$

\medskip

\noindent {\bf subcase 2b(ii)}: $\alpha \ne 0$ (without any loss of generality one can set $\alpha =1$). Then one has $p=e^a/(e^a-1)$, which corresponds to the PDE 
$$
\frac{e^{u_1}}{e^{u_1}-1}u_{12}+\frac{1}{u_3}(u_{13}-u_{23})=0.
$$
It possesses four conservation laws:
$$
\partial_1 (u_3 e^{u_2})+\partial_3 (e^{u_2-u_1}-e^{u_2}) =0,
$$
$$
\partial_1 \ln u_3+\partial_2  \ln\left(\frac{1-e^{u_1}}{u_3} \right) =0,
$$
$$
\partial_2 \left( \frac{u_3}{1-e^{u_1}}     \right) +\partial_3  \left(\ln(1-e^{u_1})-u_1 \right)=0,
$$
$$
\partial_1 \left( \frac{u_2 u_3}{2}+u_3 \right)+\partial_2 \left( \frac{u_1 u_3 \left(e^{u_1}+1\right) }{2 \left(e^{u_1}-1\right)}-u_3 \right)+\partial_3 \left( \frac{u_1^2-u_1 u_2}{2}-u_1 \ln \left(1-e^{u_1}\right) -{Li}_2\left(e^{u_1}\right)  \right)=0.
$$
Applying steps 3 and 4, we can show that semi-discrete versions of the first three conservation laws correspond to the differential-difference equation
\begin{equation}
(1-e^{(T_{12}u-T_{2}u)/ \epsilon})T_1 u_3  = (1-e^{(T_{1}u-u)/ \epsilon})T_2 u_3,
\label{dd5}
\end{equation}
which possesses the Lax pair
$$
T_{1} \psi=(1-e^{(T_1 u- u)/\epsilon}) T_{2}\psi-e^{(T_1 u- u)/\epsilon} \psi, ~~~~
\epsilon \psi_3=u_3 T_{2}\psi+u_3 \psi.
$$
Setting $v=e^{u/ \epsilon}$ we can rewrite equation \eqref{dd5} in the form
$$v (T_{12}v-T_2 v)T_1 v_3=T_1 v (T_1 v-v)T_2 v_3.$$ 

\medskip

\noindent {\bf subcase 2c}: both $p$ and $ q$ are non-constant. Subtracting the ODEs for $p$ and $q$ from each other and separating the variables gives $p'=\alpha-p^2, \ q'=\alpha-q^2$. There are further subcases depending on the value of the integration constant $\alpha$.

\medskip

\noindent {\bf subcase 2c(i)}: $\alpha=0$. Then one can take $p=1/a, \ q=1/b$, which results in the  PDE
$$
\left(\frac{1}{u_1}-\frac{1}{u_2}\right)u_{12}+\frac{1}{u_3}(u_{13}-u_{23})=0.
$$
It possesses four conservation laws:
$$
\partial_1 \ln \left(\frac{u_2}{u_3} \right)+\partial_2 \ln \left(\frac{u_3}{u_1} \right) =0,
$$
$$
\partial_1 \left( \frac{u_3}{u_2}     \right) +\partial_3  \ln \left(1-\frac{u_1}{u_2} \right)=0,
$$
$$
\partial_2 \left( \frac{u_3}{u_1}     \right) +\partial_3  \ln \left(1-\frac{u_2}{u_1} \right)=0,
$$
$$
\partial_1 \left( u_2^2 u_3 \right)-\partial_2 \left( u_1^2 u_3 \right)+\partial_3 \left( u_1^2 u_2-u_2^2 u_1   \right)=0.
$$
Applying steps 3 and 4, we can show that semi-discrete versions of the first three conservation laws correspond to the differential-difference equation
\begin{equation}
(T_2 \triangle_1u)(\triangle_2u) T_1 u_3  = (T_1 \triangle_2u)(\triangle_1u) T_2 u_3 ,
\label{dd6}
\end{equation}
which appeared in \cite{BK99}. Equation \eqref{dd6} possesses the Lax pair
$$
T_{1} \psi=\frac{\triangle_1 u}{\triangle_2 u} T_{2}\psi+\left(1-\frac{\triangle_1 u}{\triangle_2 u} \right) \psi, ~~~~
\epsilon \psi_3=\frac{u_3}{\triangle_2 u}  (T_{2}\psi- \psi).
$$

\medskip

\noindent {\bf subcase 2c(ii)}: $\alpha\ne 0$ (we will consider the hyperbolic case $\alpha=1$; the trigonometric  case $\alpha=-1$ is similar). Then one can take $p=\coth a, \ q=\coth b$, which results in the  PDE
$$
\left(\coth{u_1}-\coth{u_2}\right)u_{12}+\frac{1}{u_3}(u_{13}-u_{23})=0.
$$
It possesses four conservation laws:
$$
\partial_1 \ln \left( \frac{\sinh u_2}{ u_3}    \right) -\partial_2  \ln \left(\frac{\sinh u_1}{u_3} \right)=0,
$$
$$
\partial_1 \left(u_3 \coth u_2 \right)+\partial_2 \ln \left(\frac{\sinh(u_1-u_2)}{\sinh u_2} \right) =0,
$$
$$
\partial_2 \left(u_3 \coth u_1 \right) +\partial_3  \ln \left(\frac{\sinh(u_1-u_2)}{\sinh u_1} \right)=0,
$$
\begin{gather*}
\partial_1 \left(4 u_3(1-\coth u_2-u_2\coth u_2)\right)+\partial_2(4 u_1 u_3 \coth u_1)+\\
\partial_3 \left(4 u_1-2 u_2^2-12 u_2+2(u_1-u_2-2)\ln \left(1-e^{2 u_1-2 u_2}\right)+2(u_1-u_2) \ln \left(1-e^{2 u_2-2 u_1}\right)+ \right. \\
\left. 4(u_2+1)\ln \left(1-e^{2 u_2}\right)-2 u_1 \ln \left((1-e^{-2 u_1})(1-e^{2 u_1})\right)+{Li}_2\left(e^{-2 u_1}\right)-{Li}_2\left(e^{2 u_1}\right)+ \right.\\
\left. {Li}_2\left(e^{2 u_1-2 u_2}\right)+2 {Li}_2\left(e^{2 u_2}\right)-{Li}_2\left(e^{2 u_2-2 u_1}\right)\right)=0. 
\end{gather*}
Applying steps 3 and 4, we can show that semi-discrete versions of the first three conservation laws correspond to the differential-difference equation
\begin{equation}
(T_2 \sinh \triangle_1u) (\sinh \triangle_2u) T_1 u_3  =(T_1 \sinh \triangle_2u) (\sinh \triangle_1u) T_2 u_3,
\label{dd7}
\end{equation}
which possesses the following Lax pair:
$$
T_{2} \psi=\frac{e^{2 \triangle_2 u}-1}{e^{2 \triangle_1 u}-1} T_{1}\psi+\frac{e^{2 \triangle_1 u}-e^{2 \triangle_2 u}}{e^{2 \triangle_1 u}-1}  \psi, ~~~~
\epsilon \psi_3=\frac{2 u_3}{e^{2 \triangle_1 u}-1}  (T_{1}\psi- \psi).
$$
This finishes the proof of Theorem 2.

\subsection { One discrete and two continuous variables.}

 One can show that there exist no non-degenerate integrable equations of the form
$$
\triangle_1 f+\partial_2 g+\partial_3 h=0,
$$
where $f, g, h$ are functions of $\triangle_1u, u_2, u_3$.

\section {Discrete second order quasilinear equations in 3D}

Here we present the result of classification of integrable equations of the form
$$
\sum_{i,j=1}^3 f_{ij}(\triangle u)\triangle_{ij}u=0,
$$
where $f_{ij}$ are functions of $\triangle_1u, \triangle_2u, \triangle_3u$ only.  These equations can be viewed as discretisations  of second order quasilinear  PDEs 
$$
\sum_{i,j=1}^3 f_{ij}(u_k) u_{ij}=0,
$$
whose integrability was investigated in \cite{BFT}. 

\begin{theorem} There exists a unique non-degenerate discrete second order quasilinear equation in 3D, known as   the lattice KP equation:
$$
(\triangle_1u-\triangle_2u)\triangle_{12}u+
(\triangle_3u-\triangle_1u)\triangle_{13}u+
(\triangle_2u-\triangle_3u)\triangle_{23}u=0.
$$
\end{theorem}
In different contexts and equivalent forms, it has appeared in  \cite{Date, Nijhoff84, Nijhoff86}. The proof is similar to that of Theorem 1,  and will be omitted.

\bigskip

\subsection{Two discrete and one continuous variables} 

The classification of semi-discrete integrable equations of the form
$$
f_{11}\triangle_{11}u+f_{12}\triangle_{12}u+f_{22}\triangle_{22}u+f_{13}\triangle_{1}u_3+f_{23}\triangle_{2}u_3+f_{33}u_{33}=0, 
$$
where the coefficients $f_{ij}$ are functions of $(\triangle_1u, \triangle_2u, u_3)$, gives the following result:

\begin{theorem} There exists a unique non-degenerate  second order  equations  of the above type, known  as the semi-discrete Toda lattice,
$$
(\triangle_1u-\triangle_2u)\triangle_{12}u-\triangle_1u_3+\triangle_2u_3=0.
$$
\end{theorem}
It has appeared before in  {\cite{Adler, Santini}. Again, we skip the details of calculations.

\bigskip

\subsection{One discrete and two continuous variables}

 One can show that there exist no non-degenerate semi-discrete integrable equations of the form
$$
f_{11}\triangle_{11}u+f_{12}\triangle_{1}u_2+f_{22}u_{22}+f_{13}\triangle_{1}u_3+f_{23}u_{23}+f_{33}u_{33}=0, 
$$
where the coefficients $f_{ij}$ are functions of $(\triangle_1u, u_2, u_3)$.

\section {Appendix: $\triangle$-forms of  Hirota-type  difference equations} \label{Appendix}

Below we list $\triangle$-forms of various 3D discrete integrable equations which have been discussed in the literature. The advantage of this representation is that the corresponding dispersionless limits become more clearly seen.  Although these equations have appeared under different names, most of them are  related via various gauge/Miura/B\"acklund type transformations. We have verified that all of the equations below inherit hydrodynamic reductions of their dispersionless limits, at least to the order $\epsilon^2$.

\medskip

\noindent {\bf Hirota equation} \cite{Hirota81}:
$$
\alpha T_1\tau T_{\bar 1}\tau+ \beta T_2\tau T_{\bar 2}\tau +\gamma T_3\tau T_{\bar 3}\tau =0.
$$
Dividing  by $\tau^2$ and setting $\tau=e^{u/\epsilon^2}$  we can rewrite it in the form
$$
\alpha  e^{\triangle_{1\bar 1}u}+\beta  e^{\triangle_{2\bar 2}u}+\gamma  e^{\triangle_{3\bar 3}u}=0.
$$
Its dispersionless limit is
$$
\alpha  e^{u_{11}}+\beta  e^{u_{22}}+\gamma  e^{u_{33}}=0.
$$

\medskip

\noindent {\bf  Hirota-Miwa equation} \cite{Miwa82}:
$$
\alpha T_1\tau T_{23}\tau+ \beta T_2\tau T_{13}\tau +\gamma T_3\tau T_{12}\tau =0.
$$
Dividing  by $T_1\tau T_2\tau T_3\tau/\tau$ and setting $\tau=e^{u/\epsilon^2}$ we can rewrite it in the form
$$
\alpha  e^{\triangle_{23}u}+\beta  e^{\triangle_{13}u}+\gamma  e^{\triangle_{12}u}=0.
$$
Its dispersionless limit is
$$
\alpha  e^{u_{23}}+\beta  e^{u_{13}}+\gamma  e^{u_{12}}=0.
$$

\medskip

\noindent {\bf  Gauge-invariant Hirota equation, or Y-system} \cite{Zabrodin97, Kuniba11}:

$$
\frac{T_2uT_{\bar 2}u}{T_1yT_{\bar 1}u}=\frac{(1+T_3u)(1+T_{\bar 3}u)}{(1+T_1u)(1+T_{\bar 1}u)}.
$$
 Taking log of both sides we obtain
$$
(\triangle_{2 \bar 2}-\triangle_{1\bar 1})\ln u=(\triangle_{3 \bar 3}-\triangle_{1 \bar 1})\ln (1+u).
$$
Its dispersionless limit is
$$
(\partial_2^2-\partial_1^2)\ln u=(\partial_3^2-\partial_1^2)\ln (1+u).
$$

\medskip

\noindent {\bf  Lattice KP equation}  \cite{Date, Nijhoff84, Nijhoff86}:
$$
(T_1u-T_2u)T_{12}u+(T_3u-T_1u)T_{13}u+(T_2u-T_3u)T_{23}u=0.
$$
In equivalent form,
$$
(\triangle_1u-\triangle_2u)\triangle_{12}u+
(\triangle_3u-\triangle_1u)\triangle_{13}u+
(\triangle_2u-\triangle_3u)\triangle_{23}u=0.
$$
 Its dispersionless limit is
$$
(u_1-u_2)u_{12}+(u_3-u_1)u_{13}+(u_2-u_3)u_{23}=0.
$$

\medskip

\noindent {\bf Schwarzian KP  equation} \cite{Nijhoff84, DN91, BK98, BK99, KonS02}:
$$
(T_2\triangle_1u)(T_3\triangle_2u)(T_1\triangle_3u)=(T_2\triangle_3u)(T_3\triangle_1u)(T_1\triangle_2u).
$$
Taking log of both sides we obtain
$$
\triangle_1\left(\ln \frac{\triangle_3u}{\triangle_2u}\right)+\triangle_2\left(\ln \frac{\triangle_1u}{\triangle_3u}\right)+\triangle_3\left(\ln \frac{\triangle_2u}{\triangle_1u}\right)=0.
$$
Its dispersionless limit is
$$
u_3(u_2-u_1)u_{12}+u_2(u_1-u_3)u_{13}+u_1(u_3-u_2)u_{23}=0.
$$

\medskip

\noindent {\bf Lattice spin  equation} \cite{Nijhoff90}:
$$
\left(\frac{T_{12}\tau}{T_2\tau}-1  \right)\left(\frac{T_{13}\tau}{T_1\tau}-1  \right)\left(\frac{T_{23}\tau}{T_3\tau}-1  \right)=
\left(\frac{T_{12}\tau}{T_1\tau}-1  \right)\left(\frac{T_{13}\tau}{T_3\tau}-1  \right)\left(\frac{T_{23}\tau}{T_2\tau}-1  \right).
$$
On multiplication by $T_1\tau T_2\tau T_3\tau$ it reduces to the Schwarzian KP equation. However, different representation can be obtained by taking log of both sides and setting $\tau=e^{u/\epsilon}$. This gives
$$
\triangle_1\ln \frac{e^{\triangle_3u}-1}{e^{\triangle_2u}-1}+\triangle_2\ln \frac{e^{\triangle_1u}-1}{e^{\triangle_3u}-1}+\triangle_3\ln \frac{e^{\triangle_2u}-1}{e^{\triangle_1u}-1}=0.
$$
Its dispersionless limit is
$$
\frac{e^{u_2}-e^{u_1}}{(e^{u_1}-1)(e^{u_2}-1)}u_{12}+
\frac{e^{u_1}-e^{u_3}}{(e^{u_1}-1)(e^{u_3}-1)}u_{13}+
\frac{e^{u_3}-e^{u_2}}{(e^{u_2}-1)(e^{u_3}-1)}u_{23}=0.
$$

\bigskip

\noindent {\bf Sine-Gordon equation} \cite{KonS02}:
$$
(T_2\sin \triangle_1u)(T_3\sin \triangle_2u)(T_1\sin \triangle_3u)=(T_2\sin \triangle_3u)(T_3\sin \triangle_1u)(T_1\sin \triangle_2u).
$$
Taking log of both sides we obtain
$$
\triangle_1\left(\ln \frac{\sin \triangle_3u}{\sin \triangle_2u}\right)+\triangle_2\left(\ln \frac{\sin \triangle_1u}{\sin \triangle_3u}\right)+\triangle_3\left(\ln \frac{\sin \triangle_2u}{\sin \triangle_1u}\right)=0.
$$
Its dispersionless limit is
$$
(\cot u_2-\cot u_1)u_{12}+(\cot u_1-\cot u_3)u_{13}+(\cot u_3-\cot u_2)u_{23}=0.
$$
This example is nothing but trigonometric version of the lattice spin equation.

\medskip

\noindent {\bf  Lattice mKP equation} \cite{Nijhoff90}:
$$
\frac{T_{13}\tau-T_{12}\tau}{T_1\tau}+\frac{T_{12}\tau-T_{23}\tau}{T_2\tau}+\frac{T_{23}\tau-T_{13}\tau}{T_3\tau}=0.
$$
Setting $\tau=e^{u/\epsilon}$ we obtain
$$
\triangle_1(e^{\triangle_3u}-e^{\triangle_2u})+\triangle_2(e^{\triangle_1u}-e^{\triangle_3u})+\triangle_3(e^{\triangle_2u}-e^{\triangle_1u})=0,
$$
its dispersionless limit is
$$
(e^{u_1}-e^{u_2})u_{12}+(e^{u_3}-e^{u_1})u_{13}+(e^{u_2}-e^{u_3})u_{23}=0.
$$

\medskip

\noindent {\bf  Toda equation} \cite{Zabrodin97}:
$$
\alpha T_1\tau T_2\tau +\beta \tau T_{12} \tau+\gamma T_{1\bar 3} \tau T_{23} \tau=0.
$$
Dividing by $T_1\tau T_2\tau$ and setting $\tau =e^{u/\epsilon^2}$ we get
$$
\alpha+\beta e^{\triangle_{12}u} + \gamma e^{\triangle_{23}u-\triangle_{1\bar 3}u+\triangle_{3\bar 3}u}=0,
$$
its dispersionless limit is
$$
\alpha+\beta e^{u_{12}}+\gamma e^{u_{23}-u_{13}+u_{33}}=0.
$$

\medskip

\noindent {\bf  Lattice Toda equation} \cite{Nijhoff90}:
$$
(T_1-T_3)\frac{T_2\tau}{\tau}=(T_2-T_3)\frac{T_1\tau}{\tau}.
$$
Setting $\tau =e^{u/\epsilon}$ we get
$$
\triangle_1(e^{\triangle_2u})-\triangle_2(e^{\triangle_1u})+\triangle_3(e^{\triangle_1u}-e^{\triangle_2u})=0,
$$
its dispersionless limit is
$$
(e^{u_2}-e^{u_1})u_{12}+e^{u_1}u_{13}-e^{u_2}u_{23}=0.
$$

\medskip

\noindent {\bf  Lattice mToda equation} \cite{Nijhoff90}:
$$
\left(\frac{T_{13}\tau}{T_1\tau}-1  \right)\left(\frac{T_{23}\tau}{T_3\tau}-1  \right)=
\left(\frac{T_{12}\tau}{T_1\tau}-1  \right)\left(\frac{T_{23}\tau}{T_2\tau}-1  \right).
$$
 Taking log of both sides and setting $\tau=e^{u/\epsilon}$ we get
$$
\triangle_1\ln \frac{e^{\triangle_3u}-1}{e^{\triangle_2u}-1}-\triangle_2\ln (e^{\triangle_3u}-1)+\triangle_3\ln (e^{\triangle_2u}-1)=0.
$$
Its dispersionless limit is
$$
-\frac{e^{u_2}}{e^{u_2}-1}u_{12}+
\frac{e^{u_3}}{e^{u_3}-1}u_{13}+
\frac{e^{u_3}-e^{u_2}}{(e^{u_2}-1)(e^{u_3}-1)}u_{23}=0.
$$

\bigskip

\noindent {\bf Toda equation for rotation coefficients} \cite{Doliwa00}:
$$
(T_2-1)\frac{T_1\tau}{\tau}=T_1\frac{T_2\tau}{T_{\bar 3}\tau}-\frac{T_{23}\tau}{\tau}.
$$
This equation appeared in the theory of Laplace transformations of discrete quadrilateral nets. Setting $\tau=e^{u/\epsilon}$ we obtain
$$
\triangle_2(e^{\triangle_1u})=(\triangle_1-\triangle_3)e^{\triangle_2u+\triangle_{\bar 3}u}.
$$
Its dispersionless limit is
$$
e^{u_1}u_{12}=e^{u_2+u_3}(u_{12}+u_{13}-u_{23}-u_{33}).
$$

\bigskip

\noindent {\bf One more version of the Toda equation} \cite{BK98}:
$$
T_{\bar 1 3}\tau+\alpha T_2\tau=T_{\bar 1}\tau T_3\tau \left(\frac{1}{\tau}+\alpha \frac{1}{T_{\bar 1\bar 2 3}\tau }\right).
$$
Setting $\tau=e^{-u/\epsilon}$ we obtain
$$
\triangle_3e^{\triangle_{\bar 1}u}=\alpha (\epsilon \triangle_{\bar 1 \bar 2}-\triangle_{\bar 1}-\triangle_{\bar 2})e^{\triangle_3u-\triangle_2u}.
$$
Its dispersionless limit is
$$
e^{u_1}u_{13}+\alpha e^{u_3-u_2}(u_{13}+u_{23}-u_{12}-u_{22})=0.
$$

\bigskip

\noindent {\bf Schwarzian Toda equation} \cite{BK98, BK99}:
$$
(T_1\triangle_3 u)(T_2(\triangle_1+\triangle_{\bar 2})u)(T_3\triangle_{\bar 2}u)=
(\triangle_3 u)(T_3(\triangle_1+\triangle_{\bar 2})u)(T_1\triangle_{ 2}u).
$$
Taking log of both sides we obtain
$$
\triangle_1\ln \triangle_3u+(\triangle_2-\triangle_3)\ln (\triangle_1+\triangle_{\bar 2})u+\triangle_3\ln \triangle_{\bar 2}u-\triangle_1\ln \triangle_2u+\frac{1}{\epsilon} \ln\left(1-\epsilon \frac{\triangle_{2\bar 2}u}{\triangle_2u}\right)=0.
$$
Its dispersionless limit is
$$
\frac{u_2}{u_1u_3}(u_1+u_2-u_3)u_{13}-u_{12}-u_{22}+u_{23}=0.
$$

\bigskip

\noindent {\bf BKP equation in Miwa form} \cite{Miwa82, NS97}:
$$
\alpha T_1\tau T_{23}\tau+ \beta T_2\tau T_{13}\tau+\gamma T_3\tau T_{12}\tau+\delta \tau T_{123}\tau=0.
$$
This equation can be interpreted as the permutability theorem of Moutard transformations \cite{NS97}.
Dividing  by $T_1\tau T_2\tau T_3\tau/\tau$ and setting $\tau=e^{u/\epsilon^2}$ we get
$$
\alpha  e^{\triangle_{23}u}+\beta e^{\triangle_{13}u}+\gamma  e^{\triangle_{12}u}+\delta \ e^{\epsilon \triangle_{123}u+ \triangle_{23}u
+ \triangle_{13}u+ \triangle_{12}u}=0. 
$$
Its dispersionless limit is
$$
\alpha  e^{u_{23}}+\beta  e^{u_{13}}+\gamma  e^{u_{12}}+ \delta  e^{u_{23}+u_{13}+u_{12}}=0.
$$

\bigskip

\noindent {\bf BKP equation in Hirota form} \cite{Miwa82}:
$$
\alpha T_1\tau T_{\bar 1}\tau+ \beta T_2\tau T_{\bar 2}\tau +\gamma T_3\tau T_{\bar 3}\tau + \delta T_{123} \tau T_{\bar 1 \bar 2 \bar 3} \tau=0.
$$
Dividing by $\tau^2$ and setting $\tau=e^{u/\epsilon^2}$ we get
$$
\alpha  e^{\triangle_{1\bar 1}u}+\beta  e^{\triangle_{2\bar 2}u}+\gamma  e^{\triangle_{3\bar 3}u}+ \delta  e^{\epsilon(\triangle_{123}u-\triangle_{\bar 1 \bar 2 \bar 3}u)+S}=0,
$$
where
$$
\begin{array}{c}
S=(\triangle_{1\bar 1}u+\triangle_{2\bar 2}u+\triangle_{3\bar 3}u)+
(\triangle_{12}u+\triangle_{\bar 1 \bar 2}u)+
(\triangle_{13}u+\triangle_{\bar 1 \bar 3}u)+
(\triangle_{23}u+\triangle_{\bar 2 \bar 3}u).
\end{array}
$$
Its dispersionless limit is  
$$
\alpha  e^{u_{11}}+\beta e^{u_{22}}+\gamma  e^{u_{33}}+\delta  e^{u_{11}+u_{22}+u_{33}+2u_{12}+2u_{13}+2u_{23}}=0.
$$

\bigskip

\noindent {\bf Schwarzian BKP equation} \cite{NS98, Kon-S02, TW1}:
$$
\frac{(T_1u-T_2u)(T_{123}u-T_3u)}{(T_2u-T_3u)(T_{123}u-T_1u)}=
\frac{(T_{13}u-T_{23}u)(T_{12}u-u)}{(T_{12}u-T_{13}u)(T_{23}u-u)}.
$$
 Taking log of both sides  we get
 $$
 \triangle_3\ln \frac{\epsilon \triangle_{12}u+\triangle_1u+\triangle_2u}{\triangle_1u-\triangle_2u}=
 \triangle_1\ln \frac{\epsilon \triangle_{23}u+\triangle_2u+\triangle_3u}{\triangle_3u-\triangle_2u}.
$$
Its dispersionless limit is \cite{BK05}:  
$$
u_3(u_2^2-u_1^2)u_{12}+u_2(u_1^2-u_3^2)u_{13}+u_1(u_3^2-u_2^2)u_{23}=0.
$$
It was shown in \cite{TW1} that the Schwarzian BKP equation is the only non-linearizable affine linear discrete equation consistent around a 4D cube.

\bigskip

\noindent {\bf  BKP version of the sine-Gordon equation} \cite{NS98, Kon-S02}:
$$
\frac{\sin(T_1u-T_2u)\sin(T_{123}u-T_3u)}{\sin(T_2u-T_2u)\sin(T_{123}u-T_1u)}=
\frac{\sin(T_{13}u-T_{23}u)\sin(T_{12}u-u)}{\sin(T_{12}u-T_{13}u)\sin(T_{23}u-u)}.
$$
 Taking log of both sides  we get
 $$
 \triangle_3\ln \frac{\sin(\epsilon \triangle_{12}u+\triangle_1u+\triangle_2u)}{\sin (\triangle_1u-\triangle_2u)}=
 \triangle_1\ln \frac{\sin(\epsilon \triangle_{23}u+\triangle_2u+\triangle_3u)}{\sin(\triangle_3u-\triangle_2u)}.
$$
Its dispersionless limit is  
$$
\begin{array}{c}
\sin 2u_3(\sin^2u_2-\sin^2u_1)u_{12}+\sin 2u_2(\sin^2u_1-\sin^2u_3)u_{13}+\\
\sin 2u_1(\sin^2u_3-\sin^2u_2)u_{23}=0.
\end{array}
$$

\bigskip

\noindent {\bf CKP equation} \cite{Schief03}:
$$
\begin{array}{c}
(\tau T_{123}\tau-T_1\tau T_{23}\tau-T_2\tau T_{13}\tau-T_3\tau T_{12}\tau)^2=\\
4(T_1\tau T_2\tau T_{13}\tau T_{23}\tau +T_2\tau T_3\tau T_{12}\tau T_{13}\tau +T_1\tau T_3\tau T_{12}\tau T_{23}\tau
-T_1\tau T_2\tau T_{3}\tau T_{123}\tau -\tau T_{12}\tau T_{13}\tau T_{23}\tau ).
\end{array}
$$
Multiplying by $[\tau/(T_1\tau T_2\tau T_{3}\tau)]^2$ and setting $\tau=e^{u/\epsilon^2}$ we obtain
$$
\begin{array}{c}
(e^{\epsilon \triangle_{123}u+\triangle_{23}u+\triangle_{13}u+\triangle_{12}u}-e^{\triangle_{23}u}-e^{\triangle_{13}u}-e^{\triangle_{12}u})^2=\\
4(e^{\triangle_{13}u+\triangle_{23}u}+e^{\triangle_{12}u+\triangle_{13}u}+e^{\triangle_{12}u+\triangle_{23}u}
-e^{\epsilon \triangle_{123}u+\triangle_{23}u+\triangle_{13}u+\triangle_{12}u}-
e^{\triangle_{23}u+\triangle_{13}u+\triangle_{12}u}).
\end{array}
$$
Its dispersionless limit is
$$
(e^{u_{23}+u_{13}+u_{12}}-e^{u_{23}}-e^{u_{13}}-e^{u_{12}})^2=
4(e^{u_{13}+u_{23}}+e^{u_{12}+u_{13}}+e^{u_{12}+u_{23}}-2e^{u_{23}+u_{13}+u_{12}}).
$$
It is remarkable that this dispersionless  equation decouples into the product of four dispersionless BKP-type equations: setting $u=2v$ we obtain
$$
(e^{v_{23}+v_{13}+v_{12}}+e^{v_{23}}+e^{v_{13}}+e^{v_{12}})(e^{v_{23}+v_{13}+v_{12}}-e^{v_{23}}-e^{v_{13}}+e^{v_{12}})\times
$$
$$
(e^{v_{23}+v_{13}+v_{12}}-e^{v_{23}}+e^{v_{13}}-e^{v_{12}})(e^{v_{23}+v_{13}+v_{12}}+e^{v_{23}}-e^{v_{13}}-e^{v_{12}})=0.
$$
One can show that hydrodynamic reductions of each BKP-branch of the dispersionless equation are inherited by the full CKP equation.
Multidimensional consistency of the CKP equation, interpreted as the Cayley hyperdeterminant, was established in \cite{TW2, Doliwa10}. An alternative form of the CKP equation was proposed earlier in \cite{Kashaev}.

\bigskip

\section*{Acknowledgements}

We thank Frank Nijhoff and Yuri Suris for clarifying discussions.


\begin{thebibliography}{99}



\bibitem{Adler} V.E. Adler,  The tangential map and associated integrable equations, J. Phys. A {\bf 42}, no. 33 (2009) 332004, 12 pp.

\bibitem{ABS1} V.E. Adler, A.I. Bobenko and Yu.B.  Suris, Classification of integrable equations on quad-graphs. The consistency approach, Comm. Math. Phys. {\bf 233}, no. 3 (2003) 513--543.

\bibitem{ABS12} V.E. Adler, A.I. Bobenko and Yu.B. Suris,  Classification of integrable discrete equations of octahedron type, Int. Math. Res. Not. IMRN  no. 8 (2012) 1822--1889.

\bibitem{Suris} R. Ball, M. Petrera and Yu.B. Suris, What is integrability of  discrete variational systems arXiv:1307.0523v1.

\bibitem{BK98} L.V. Bogdanov and  B.G. Konopelchenko,  Analytic-bilinear approach to integrable hierarchies. II. Multicomponent KP and 2D Toda lattice hierarchies, J. Math. Phys. {\bf 39}, no. 9 (1998) 4701--4728.

\bibitem{BK99} L.V. Bogdanov and B.G. Konopelchenko,  M\"obius invariant integrable lattice equations associated with KP and 2DTL hierarchies,  Phys. Lett. A {\bf 256}, no. 1 (1999) 39--46.

\bibitem{BK05} L.V. Bogdanov and B.G. Konopelchenko, On dispersionless BKP hierarchy and its reductions, J. Nonlinear Math. Phys. {\bf 12}, suppl. 1 (2005)  64--73.

\bibitem{BFT} P.A. Burovskii, E.V. Ferapontov and S.P. Tsarev, Second order quasilinear PDEs and conformal structures in projective space, International J. Math. { \bf 21}, no. 6 (2010) 799-841.

\bibitem{Date} E. Date, M. Jimbo and T. Miwa, Journ. Phys. Soc. Japan 51 (1982) 4125.

\bibitem{Doliwa00} A. Doliwa,  Lattice geometry of the Hirota equation. SIDE III--symmetries and integrability of difference equations (Sabaudia, 1998), 93-100, CRM Proc. Lecture Notes, 25, Amer. Math. Soc., Providence, RI, (2000).

\bibitem{Doliwa10} A. Doliwa,  The C-(symmetric) quadrilateral lattice, its transformations and the algebro-geometric construction, J. Geom. Phys. {\bf 60}, no. 5 (2010) 690--707.

\bibitem{DN91} I.Ya. Dorfman and F. W. Nijhoff,  On a (2 + 1)-dimensional version of the KricheverÐNovikov equation,  Physics Letters A {\bf 157}, no. 2-3 (1991) 107--112.

\bibitem{FK} E.V. Ferapontov and K.R. Khusnutdinova, On integrability of (2+1)-dimensional quasilinear systems, Comm. Math. Phys. {\bf 248} (2004) 187-206.

\bibitem{FerM} E.V. Ferapontov and A. Moro,
Dispersive deformations of hydrodynamic reductions of 2D
dispersionless integrable systems, J. Phys. A: Math. Theor. {\bf 42}
(2009) 035211, 15pp.

\bibitem{FMN} E.V. Ferapontov, A. Moro and V.S. Novikov, Integrable equations in $2+1$ dimensions: deformations of dispersionless limits, J. Phys. A: Math. Theor. {\bf 42} (2009)  18pp.

\bibitem{FMoss} E.V. Ferapontov and J. Moss, Linearly degenerate PDEs and quadratic line complexes, arXiv:1204.2777.

\bibitem{FNR} E.V. Ferapontov, V.S. Novikov and I. Roustemoglou, Towards the classification of integrable differential-difference equations in 2 + 1 dimensions, J. Phys. A: Math. Theor. {\bf 46} (2013) 13pp.


\bibitem{FHZ} E.V. Ferapontov, B. Huard and A. Zhang,
On the central quadric ansatz: integrable models and Painleve reductions, J. Phys. A:  Math. Theor. {\bf 45}  (2012) 195204.

\bibitem{Hirota81} R. Hirota,  Discrete analogue of a generalized Toda equation, J. Phys. Soc. Japan 50 (1981) 3785--3791.

\bibitem{HN} B. Huard and V.S. Novikov, On classification of integrable Davey-Stewartson type equations, J. Phys. A {\bf 46}, no. 27 (2013) 275202, 13 pp.



\bibitem{Kashaev} R.M.  Kashaev,  On discrete three-dimensional equations associated with the local Yang-Baxter relation,  Lett. Math. Phys. {\bf 35} (1996) 389--397.

\bibitem{KingS03} A.D. King and W.K. Schief, Tetrahedra, octahedra and cubo-octahedra: integrable geometry of multi-ratios, J. Phys. A {\bf 36}, no. 3 (2003) 785--802.

\bibitem{KonS02} B.G. Konopelchenko and W.K.  Schief,  Menelaus' theorem, Clifford configurations and inversive geometry of the Schwarzian KP hierarchy, J. Phys. A {\bf 35}, no. 29 (2002) 6125--6144.

\bibitem{Kon-S02} B.G. Konopelchenko and W.K.  Schief,  Reciprocal figures, graphical statics, and inversive geometry of the Schwarzian BKP hierarchy,  Stud. Appl. Math. {\bf 109}, no. 2 (2002) 89--124.

\bibitem{Kuniba11} A. Kuniba, T. Nakanishi and J. Suzuki, T-systems and Y-systems in integrable systems,  J. Phys. A: Math. Theor. {\bf 44} (2011) 103001 (146pp).

\bibitem{Santini} D. Levi, L.  Pilloni and P.M. Santini,  Integrable three-dimensional lattices, J. Phys. A {\bf 14}, no. 7 (1981) 1567--1575.

\bibitem{Lobb}   S. Lobb and F. Nijhoff,  Lagrangian multiforms and multidimensional consistency, J. Phys. A {\bf 42}, no. 45 (2009) 454013, 18 pp. 

\bibitem{Lobb09} S.B. Lobb,  F.W. Nijhoff and G.R.W. Quispel, Lagrangian multiform structure for the lattice KP system,  Journal of Physics A {\bf 42}, no. 47 (2009)  472002 (11 pp).


\bibitem{Miwa82} T. Miwa, On Hirota's difference equation, Proc. Japan Acad. {\bf 58} ser. A (1982) 9--12.

\bibitem{Nijhoff86} F.W. Nijhoff, The direct linearizing transform for three-dimensional lattice equations,
Physica {\bf 18}D (1986) 380--381.

\bibitem{Nijhoff84} F.W. Nijhoff, H.W.  Capel, G.L. Wiersma and G.R.W.  Quispel,  B\"acklund transformations and three-dimensional lattice equations, Phys. Lett. A {\bf 105}, no. 6 (1984) 267--272.

\bibitem{Nijhoff90} F.W. Nijhoff and H.W. Capel, The direct linearization approach to hierarchies of integrable PDEÕs in 2+1 dimensions. I. Lattice equations and the differential-difference hierarchies, Inverse Problems {\bf 6}, no. 4 (1990) 567--590.


\bibitem{NS97} J.J.C. Nimmo and W.K. Schief, Superposition principles associated with the Moutard transformation: an integrable discretization of a (2+1)-dimensional sine-Gordon system,
 Proc. R. Soc. Lond. A {\bf 453} (1997) 255-279, doi: 10.1098/rspa.1997.0015.
 
\bibitem{NS98} J.J.C. Nimmo and W.K. Schief, An integrable discretization of a (2+1)-dimensional sine-Gordon equation, Stud. Appl. Math. {\bf 100}, no. 3 (1998) 295--309.






\bibitem{Schief03} W.K. Schief,  Lattice geometry of the discrete Darboux, KP, BKP and CKP equations. Menelaus' and Carnot's theorems, J. Nonlinear Math. Phys. {\bf 10}, supp. 2 (2003) 194--208.




\bibitem{TT} K. Takasaki and T. Takebe,  Integrable hierarchies and dispersionless limit, Rev. Math. Phys. {\bf 7}, no. 5 (1995) 743--808.

\bibitem{TW1} S.P. Tsarev and T.  Wolf,  Classification of three-dimensional integrable scalar discrete equations. Lett. Math. Phys. {\bf 84}, no. 1 (2008) 31--39.

\bibitem{TW2} S.P. Tsarev and T.  Wolf,  Hyperdeterminants as integrable discrete systems,  J. Phys. A {\bf 42}, no. 45 (2009) 454023, 9 pp. 

\bibitem{Wiersma1} G.L. Wiersma and H.W.  Capel,  Lattice equations, hierarchies and Hamiltonian structures: the Kadomtsev-Petviashvili equation, Phys. Lett. A {\bf 124}, no. 3 (1987) 124--130.

\bibitem{Wiersma2} G.L. Wiersma and H.W. Capel,  Lattice equations, hierarchies and Hamiltonian structures. II. KP-type of hierarchies on 2D lattices, Phys. A 149 (1988), no. 1-2, 49Ð74.

\bibitem{Wiersma3} G.L. Wiersma and H.W. Capel, Lattice equations, hierarchies and Hamiltonian structures. III. The 2D Toda and KP hierarchies, Phys. A 149 (1988), no. 1-2, 75--106.

\bibitem{Zabrodin97} A. Zabrodin, A survey of Hirota's difference equations, Theor. Math. Phys. {\bf 113} (1997) 1347--1392.


\end{thebibliography}
\end{document}